\documentclass[sn-mathphys]{sn-jnl}
\pdfoutput=1

\usepackage[table]{colortbl}

\usepackage{cleveref}
\usepackage{geometry}
\usepackage{amsfonts}
\usepackage{amsmath}
\usepackage{graphicx}
\usepackage{array}
\usepackage{float}
\usepackage{color, colortbl}
\usepackage[labelfont=bf]{caption}
\usepackage{subcaption}
\definecolor{Gray}{gray}{0.6}
\usepackage{multirow}
\usepackage{tabularx}
\usepackage{algorithm}

\usepackage{etoolbox}
    \makeatletter
    \patchcmd{\ps@titlepage}
    {\hbox to \hsize{\hfill Springer Nature 2021 \LaTeX\ template\hfill}} %
    {\hbox to \hsize{\hfill \hfill}} 
    {}
    {}
    \patchcmd{\ps@headings}
    {\hbox to \hsize{\hfill Springer Nature 2021 \LaTeX\ template\hfill}} %
    {\hbox to \hsize{\hfill \hfill}} 
    {}
    {}
    \patchcmd{\ps@headings}
    {\hbox to \hsize{\hfill Springer Nature 2021 \LaTeX\ template\hfill}} %
    {\hbox to \hsize{\hfill \hfill}} 
    {}
    {}
    \makeatother

\jyear{2022}%

\theoremstyle{thmstyleone}%
\theoremstyle{thmstyletwo}%

\theoremstyle{thmstylethree}%

\begin{document}

\title[Inhibition's role in network resilience]{A high-efficiency model indicating the role of inhibition in the resilience of neuronal networks to damage resulting from traumatic injury}


\author*[1]{\fnm{Brian L.} \sur{Frost}}\email{b.frost@columbia.edu}

\author[2]{\fnm{Stanislav M.} \sur{Mintchev}}\email{mintchev@cooper.edu}

\affil[1]{\orgdiv{Electrical Engineering}, \orgname{Columbia University}, \orgaddress{\street{500 W 120th St}, \city{New York}, \state{NY}, \postcode{10027}, \country{USA}}}

\affil[2]{\orgdiv{Mathematics}, \orgname{The Cooper Union}, \orgaddress{\street{41 Cooper Sq}, \city{New York}, \state{NY}, \postcode{10003},  \country{USA}}}


\abstract{Recent investigations of traumatic brain injuries have shown that these injuries can result in conformational changes at the level of individual neurons in the cerebral cortex. Focal axonal swelling is one consequence of such injuries and leads to a variable width along the cell axon. Simulations of the electrical properties of axons impacted in such a way show that this damage may have a nonlinear deleterious effect on spike-encoded signal transmission. The computational cost of these simulations complicates the investigation of the effects of such damage at a network level. We have developed an efficient algorithm that faithfully reproduces the spike train filtering properties seen in physical simulations. We use this algorithm to explore the impact of focal axonal swelling on small networks of integrate and fire neurons. We explore also the effects of architecture modifications to networks impacted in this manner. In all tested networks, our results indicate that the addition of presynaptic inhibitory neurons either increases or leaves unchanged the fidelity of the network's processing properties with respect to this damage.}




\maketitle

\section{Introduction}

{\it Focal axonal swelling} is an abnormality that sometimes develops in neurons subjected to physical stress~\cite{Maia2014}.
In recent years, as medical science has taken up the goal of understanding the various mechanisms by which concussions affect the brain, progressively more accurate microscope imaging technology has shown that traumatic brain injury can lead to structural deformations in neural axon morphology within regions of affected brain tissue~\cite{Johnson2013,Tagge2018}. 
Specifically, impact trauma can cause the severance of microtubules involved in various cell transport mechanisms~\cite{Maxwell1997,Wu2021}, resulting in material pileups at points along the length of the cell axon that cause widening of the cell body~\cite{Tang-Schomer2012,Wang2011}.
As a consequence the electrical impulse transmission capabilities of the neuron are affected significantly~\cite{Maia2013}.

From a biophysical perspective, the impact of axonal swelling on electrical conduction from soma to dendrite is likely due to the fact that the compromised cell body acts as a variable-(rather than constant-)width cable. Investigations into the phenomenology arising from such altered morphology have been carried out in, e.g., \cite{Debanne2011,Maia2015,Maia2013,Manor1991,Ramon1975} 
. Specifically, \cite{Maia2013} presents a detailed mathematical study by way of numerical solution to variable-width cable PDEs, wherein the authors have observed a variety of effects on sequences of traveling wave packets traversing this sort of non-homogeneity while propagating diffusively. In particular, this study has indicated that such sequences are subject to a trichotomy of consequences including partial packet loss, total transmission blockage, and partial reflection; the specifics of each case depend on the parameter regime for the selected model.

Of the three categories of effects discussed in~\cite{Maia2013}, 
partial packet loss presents a compelling candidate for further study since it indicates that a swollen axon may behave as a non-linear frequency filter in the sense of signal processing. A systematic analysis of the nature of the packet loss using the aforementioned PDE model reveals a highly selective deletion scheme that is at least somewhat dependent on the time-sequencing of the packets. For example, there are scenarios in which a pair of closely spaced packets will transmit across the cable deformation reliably, but a triple with identical time spacing will see its third spike deleted~\cite{Maia2017}.
Examples of this nature suggest a degree of complexity that merits investigation both from the perspective of rigorous analysis at the single-axon level and from the point of view of impact on network activity and information processing.


The present article aims to address how network-level properties (both cell and connection type, as well as connectivity and strength of connections) may endow a subpopulation of neurons with greater (resp., lesser) resilience in the presence of damage due to focal axonal swelling.  Damage of the type described is modeled via decreased fidelity of the connections, which is exhibited in the form of transmuted temporal spiking sequences from presynaptic cells.   

Various studies of the consequences of concussive damage at the network level have been carried out recently; see, e.g., \cite{Sharp2014} for a thorough overview of the seminal literature. As it relates to the focus of the present discussion,~\cite{Rudy2016}
used the FORCE algorithm of Sussillo and Abbott~\cite{Sussillo2009}
to establish a measure of tolerance to damage that shows a gradual decline in network plasticity-recovery capabilities as the proportion of damaged connections is increased. Notably, this analysis relied on a firing rate model (based on assumptions regarding the effects of axonal swelling damage on the spike-emission rates of affected cells) implemented for a spiking neuronal network tasked with image processing and decision making. Subsequent work in~\cite{Lusch2018} uses a similar paradigm by leveraging a convolutional neural network approach. While illuminating, this line of work carries the drawback of potentially overlooking the specifics of neurobiology and detailed axon signal-transduction dynamics at the single-cell level. By appealing directly to and artificially imposing a value for the firing rate parameter; this {\it en gros} approach skips over the mechanisms at work within axons that have incurred physical defects, in favor of observables that measure a neuronal population's global network activity~\cite{Ermentrout2010}.
These remarks notwithstanding, a careful critical examination of the work to date on this topic reveals a significant computational challenge: to full detail, numerical investigation of this sort of network would involve intensive PDE simulations for each damaged axon in tandem with a resource-intensive neuronal network implementation. The complexity of such a study renders it prohibitive except in the case of very small networks over short time periods, a severe limitation if one has the objective of scaling to a size that would allow for modeling compromised neural tissue within, e.g., the mammalian nervous system~\cite{Herculano_Houzel2009}.

Instead of working with an {\it a priori} firing rate model, which averages over the details of dynamics at the cellular/node level, our study extracts firing rate data as an observable {\it collected from} a specific output site within a spiking network of integrate-and-fire neurons~\cite{Ger-2002}. The use of a simple one-dimensional integrate-and-fire model at the cell level allows for an efficient simulation of the spiking network that retains the basic features of voltage evolution and depolarization without slowing down the numerics~\cite{Izhikevich2004}. We aim to investigate especially the manner in which the precise voltage dynamics of individual neurons influence the overall impact of traumatic stress, on the way to understanding whether specific detailed adaptations in network architecture can mitigate this impact. To circumvent the simulation of the full axon PDE diffusion model proposed by Maia and Kutz, we implement a high-accuracy predictive machine learning algorithm that allows the recovery of a post-damage time series of spikes given precise knowledge of the input spike train. A statistical analysis shows that this algorithm is extremely accurate; in particular, sufficiently so to warrant its implementation {\it entirely instead of} the non-linear PDE model. This advance speeds up the numerics significantly. While the networks we have investigated are only of modest size, the methodology proposed herein can be adapted strategically to scale for larger networks, more sophisticated connection topology, and features at the cellular level that are ignored in simple integrate-and-fire schemes. We have proposed and partially investigated one such amenable extension in network size via the {\it layered network} formalism~\cite{Vogels2005,VogelsRajanAbbott2005,VogelsAbbott2007}.


A fundamental challenge in the present framework is the determination of a measure that serves as a reliable indicator of a network's task performance capabilities. For neurons subjected to stochastically generated spike trains, the effects of swelling-induced traumatic stress on spike-arrival-time based encoding are direct but not particularly illuminating. One assumes that a time sequence of spikes -- modeled by a binary string of 1’s and 0’s with each bit representing a fixed small time interval -- is observed after having traversed an axonal deformation. It features fewer spikes because of spike deletion, but otherwise involves no transmutation (dilation or contraction of the intervals between spikes). This results in a discretized neural code exhibiting greater numbers of consecutive 0’s, less frequently interspersed with 1’s; equivalently, a specific spike train will see an increase in the lengths of inter-spike intervals and an effective decrease in spike arrival frequency. While these bottom-line effects are clear, the stochastic variability typical in biological neurons obscures the interpretation that precise bit-string encoding is the most appropriate conduit of information on damage within the neural code~\cite{Abeles1994,Bialek1991,Haslinger2010,Lestienne1996}.

We focus instead on stimulus and response firing rate measurements averaged over many realizations. A spiking network representing a neural tissue with a specific neural processing task is bathed in a global stimulus modeled by a Poisson spike train with a tunable frequency. In the spirit of electrode recordings from a specific location \textit{in vivo}, a target neuron is selected as an output site and its spiking statistics are stored and analyzed for various frequencies of the global stimulus. This leads to a natural input-output relation between the frequency of the injected stimulus and the mean inter-spike interval at the recording site that typically features a specific monotone curve, the details of which are dependent on the characteristics of the particular network under investigation. In turn, changes to this graph indicate the effect that damage has, both at the single-cell level and on the network’s at-large signal processing capabilities. 


Proceeding in this manner, we are able to analyze successively more complex layered feedforward networks for input-output frequency response. We determine a measurement that assesses the difference between response curves of a damaged network and its undamaged analog (identical cell type of inhibitory or excitatory variety, and network architecture), and this allows us a direct view into characteristics that make a network more or less susceptible to damage from this type of traumatic injury. We show evidence indicating that network plasticity – specifically, a strategic addition of inhibitory cells or undamaged axonal connections – serves to attenuate the effects of damage, altogether pointing to mechanisms that may be at play in neural tissues as they recover from concussive trauma.

{\bf The paper is organized as follows.} In \Cref{methods} we present the details of the model investigated in the study, specifically, its properties as a network of interacting dynamical systems. We also motivate and define the measure leveraged in deducing the effects of focal axonal swelling on this type of network model. In \Cref{results} we provide a summary of results for the exhaustive study carried out for basic two-(modified to three-)element networks, as well as results from the exploratory analysis of layered networks by way of the principles developed in the small-network study. We also provide detailed discussion of the observed phenomenology. In \Cref{conclusions} we summarize the main takeaways from the results presented in \Cref{results} and provide a look ahead into possible future work in this setting.

\section{Methods} \label{methods}

\par{This investigation focuses on spiking networks of integrate-and-fire neurons employing axonal connections that can be either healthy or damaged by swelling resulting from traumatic stress. The present setting constitutes a specific example of a \textit{network of dynamical systems}, featuring a complex interplay between local intracellular dynamics and interactions between cells~\cite{Young2022}.
A mathematical formalism for studying how the system changes over time can be achieved via a triple consisting of (1) a directed graph (the vertices and edges of which indicate the network structure), (2) the local dynamical rules governing intracellular processes at each vertex, and (3) the precise communication protocol in place between vertices adjacent due to the presence of an edge, the direction of which serves to distinguish between pre- and post-synaptic cells in the sense of neuroscience. We proceed to specify here parts (2) and (3) in this formalism precisely for our model, thereby drawing attention to the lexicon that we will use to describe and interpret the results of the paper. Part (1) will be detailed after this, in preparation for presentation and discussion of the results.}

\subsection{Spiking integrate-and-fire neurons}

\par{The spiking network simulation method used here is a modified version of the synchronous or clock-driven algorithm presented by Brette et al~\cite{Brette2007}, 
to which we have introduced a refractory period for each neuron. The latter amounts to a dynamical rule ensuring that sufficient time has passed since a neuron has last spiked before it is able to spike again. This mandatory quiescent phase places a \textit{de facto} bound on individual neural firing rates; its incorporation into our model is motivated by the behavior of physical neurons~\cite{dayan2001theoretical}. 
The communication between cells in these networks is achieved via the processing of incoming spikes, which serve to affect the local dynamics by resulting in contributions to the receiving cell's voltage. An extensive treatise comparing clock-driven algorithms to their event-driven counterparts is given in~\cite{Rudolph2006}; the latter more biologically accurate alternative is not feasible in the present context.}

\par{Each neuron processes spikes according to an integrate-and-fire model much like that presented in~\cite{Ger-2002}.
A neuron can be \textit{excitatory} or \textit{inhibitory}, which means either a positive or negative contribution to the rate of change of voltage, $V$, of any neuron that is post-synaptic to it. Each neuron has three non-negative state variables that are updated at each timestep in the simulated dynamics -- a voltage $V$, and excitatory/inhibitory potentials $J_E$ and $J_I$ which determine the impact of incoming excitatory and inhibitory spikes on the voltage evolution. Assuming there are no spikes at the input of a given cell, its state variables evolve as
\begin{align}
	\tau_V \frac{dV}{dt} &= -V + J_E - J_I \label{dvdt}\\
	\tau_J \frac{dJ_E}{dt} &= -J_E\label{dJEdt}\\ 
	\tau_J \frac{dJ_I}{dt} &= -J_I\label{dJIdt}.
\end{align}}
\par{The time constants $\tau_V$ and $\tau_J$ determine the rate of these evolutions. Considering Equation \ref{dvdt}, we can see that in the absence of input, the voltage decays exponentially. Equations \ref{dJEdt} and \ref{dJIdt} provide that $J_E$ and $J_I$ also decay exponentially in the absence of spikes. Excitatory and inhibitory potentials lead to increases and decreases, respectively, to the rate of change of $V$, allowing the voltage to increase when the $J_E$ term dominates.}
\par{The input of a spike causes an instantaneous jump by $w$ in potential $J_E$ (resp., $J_I$) if the spike is produced by a presynaptic excitatory (resp., inhibitory) cell. Cells have a threshold voltage $V_T$ as well as a refractory period $T_R$ which determine when a cell is able to spike. A neuron will \textit{spike} if $V$ rises above threshold voltage once a refractory period has passed.}

\par{The routine followed at each time step, separated by $\Delta t$, of the simulation is presented in Algorithm \ref{networkloop}. In particular, note that voltage reset and spike emission occur only if voltage is above threshold \textit{after} a refractory period has elapsed, a provision that in theory allows the voltage to fluctuate above and below threshold on the refractory timescale. The numerical values of the parameters in Algorithm \ref{networkloop} used in our simulation can be found in Table~\ref{algparams}.}

\begin{algorithm}
\caption{Network loop followed at each discrete timestep}\label{networkloop}
\begin{algorithmic}[1]
\For {$\textit{cell} \textbf{ in }\textit{network}$}
\State{$\textit{cell.time} += \Delta t$}
\For {$\textit{input} \textbf{ in } \textit{cell.inputs}$}
\If {$\textit{input} == 1$}
\State $J_E = J_E +w$
\ElsIf {$\textit{input} == -1$}
\State $J_I = J_I + w$
\EndIf
\EndFor
\State $\textbf{update } \textit{cell} \text{ via the solution to the ODEs}$
\If {$\textit{cell.V} \geq V_{T} \textbf{ and } \textit{cell.time} \geq T_R$}
\State $\textit{cell.V} = 0$
\State $\textit{cell.time} = 0$
\If {$\textit{cell.type} = \textit{excitatory}$}
\State $\textit{cell.outputs} = 1$
\ElsIf {$\textit{cell.type} = \textit{inhibitory}$}
\State $\textit{cell.outputs} = -1$
\EndIf
\EndIf
\EndFor
\end{algorithmic}
\end{algorithm}

\begin{table}
\begin{center}
\begin{tabular}{c c c p{6.5cm}}
Parameter & Units & Value & Significance\\
\hline	
	$\Delta t$ & ms & 0.1 & Discrete timestep \\
	$\tau_V$ & ms & 18 & Voltage time constant \\
	$\tau_J$ & ms & 5 & $J_E$ and $J_I$ time constant \\
	$w$ & mV &  0.5 & Discrete potential step from incoming spike\\ 
	$V_T$ & mV & 0.2 & Threshold voltage\\ 
	$T_R$ & ms & 1& Refractory period
\end{tabular}
\end{center}
	\caption{Parameters used in simulating networks of neurons, as described in Algorithm \ref{networkloop}. Values were informed by~\cite{Brette2007,Ger-2002,Hansel1998}. 
	}
\label{algparams}
\end{table}

\subsection{Axonal transmission and damage} \label{methods::transmission}

\par{In the case of an undamaged connection, the spikes from one neuron are incorporated into the dynamics of connected neurons with no intermediate processing. However, if a connection is damaged, the spike train is first subjected to a filter that integrates the effects of the damage before the processing of the connected neurons. This filter can either delete spikes or allow them to be transmitted across the connection faithfully.}

\par{The damage filter implemented in this study is inspired by the model of axonal swelling presented and analyzed by Maia and Kutz~\cite{Maia2013}. This model was derived by introducing a continuous deviation of cable radius in the active cable equation given by
\begin{align}
	\frac{\partial V}{\partial x} &= \frac{D}{d(x)}\frac{\partial}{\partial x}\bigg( \frac{d^2(x)}{r_L(x)} \frac{\partial V}{\partial x} \bigg) + V(V-a)(V-1) - R\\
	\frac{\partial R}{\partial t} &= bV - cR
\end{align}
where $V$ is the voltage, $x$ is distance along the axon, $d(x)$ is the radius of the axon at position $x$, $r_L(x)$ is the resistance of the axon at position $x$, $R$ is a \textit{gating variable} (which models the probability of ion channels being open, as in the Hodgkin-Huxley model~\cite{Hodgkin1952}), and $a$, $b$, $c$ and $D$ are physical constants. The damage is implemented by letting $d(x)$ grow from width $\delta_B$ to $\delta_A$ (that is, to \textit{swell}) over a distance $\delta_T$, according to 
\begin{equation}
	d(x) =  \begin{cases} \delta_B, & \text{for } x\leq 0 \\
		\tilde{d}(x), & \text{for } 0 \leq x \leq \delta_T \\
		\delta_A, & \text{for } x\geq\delta_T \end{cases}
\end{equation}
where
\begin{equation}
    \tilde{d}(x) = \delta_B + (\delta_A - \delta_B)\bigg(10\frac{x^3}{\delta_T^3} - 15 \frac{x^4}{\delta_T^4} + 6\frac{x^5}{\delta_T^5}\bigg).
\end{equation}
The resistance in the present setting is assumed to be a constant 1 $\Omega$. The constants' physical interpretations, units, and values used to calibrate the subsequent studies can be seen in Table \ref{consts}.

\begin{table}
\begin{center}
\begin{tabular}{c c c p{6.5cm}}
Constant & Units & Value & Significance\\
\hline
$D$ & cm$^2$/$\mu$F & 0.02 & Inverse of specific membrane capacitance\\
$a$ & mV & 0.1 & Fitzhugh-Nagumo offset voltage\\
$b$ & $\mu$A$\cdot{\Omega}/\text{cm}^2$ & 0.01 & Voltage contribution to $R$ evolution\\
$c$ & Hz & 0.05& $R$ contribution to $R$ evolution \\
$d_A$ & cm & 4 & Final diameter of axon\\
$d_B$ & cm & 2 & Initial diameter of axon\\
$d_T$ & cm & 0.25 & Distance across which swelling occurs\\

\end{tabular}
\end{center}
\caption{Constants used in our simulations of the active cable model, along with their units and biological significance. Values are informed by those used by Maia and Kutz \cite{Maia2013}}
\label{consts}
\end{table}}

\par{A numerical solution of this PDE model can be carried out by an adaptation of the spectral method~\cite{Thapa2014} to nonlinear equations~\cite{Fornberg1994}. The resulting \textit{pseudo}-spectral method employed by Maia and Kutz is detailed in the Appendix. Since the solution of this system of partial differential equations is computationally expensive for a large-scale network simulation, the shortcut that we implement for the filtering amounts to a statistical algorithm trained from the behavior of the damaged active cable model. The details of this \textit{damaged axon prediction algorithm} (DAPA) are presented in the Appendix, with a side-by-side comparison with the output of the pseudo-spectral numerical method. This comparison indicates that the shortcut produces results statistically equivalent to the pseudo-spectral method.}

\subsection{Quantifying the effects of damage} \label{methods::metric}

\par{In assessing the effects of damage on a network's capability for rate-encoding, we propose a measure based on the mean inter-spike interval $I_{av}$ of a network's output over an interval of frequencies. The nonlinearity of the network model precludes studying the frequency response via tonal analysis (i.e. stimuli with one frequency component). Networks are instead stimulated by a Poisson spiketrain\cite{heeger2000poisson} presented as a uniform background excitatory stimulus to all neurons. For each input frequency $\lambda$ (Hz), at each time step, we simulate a single sample from the Bernoulli distribution with ``success" probability $p = \lambda \Delta t$. A success leads to a spike being processed as an input at every neuron. This spike is handled in the same way as a spike from a presynaptic (excitatory) neuron.}

\par{For each input stimulus frequency $\lambda$, many realizations of the input are simulated to obtain an associated $I_{av}(\lambda)$ for a given network architecture and damage paradigm. Reported $I_{av}(\lambda)$ values are averaged across all realizations; in this study, the presented average interspike intervals are averaged over 10,000 realizations.}

\par
\begin{figure}
\centering
\includegraphics[width=0.5\textwidth]{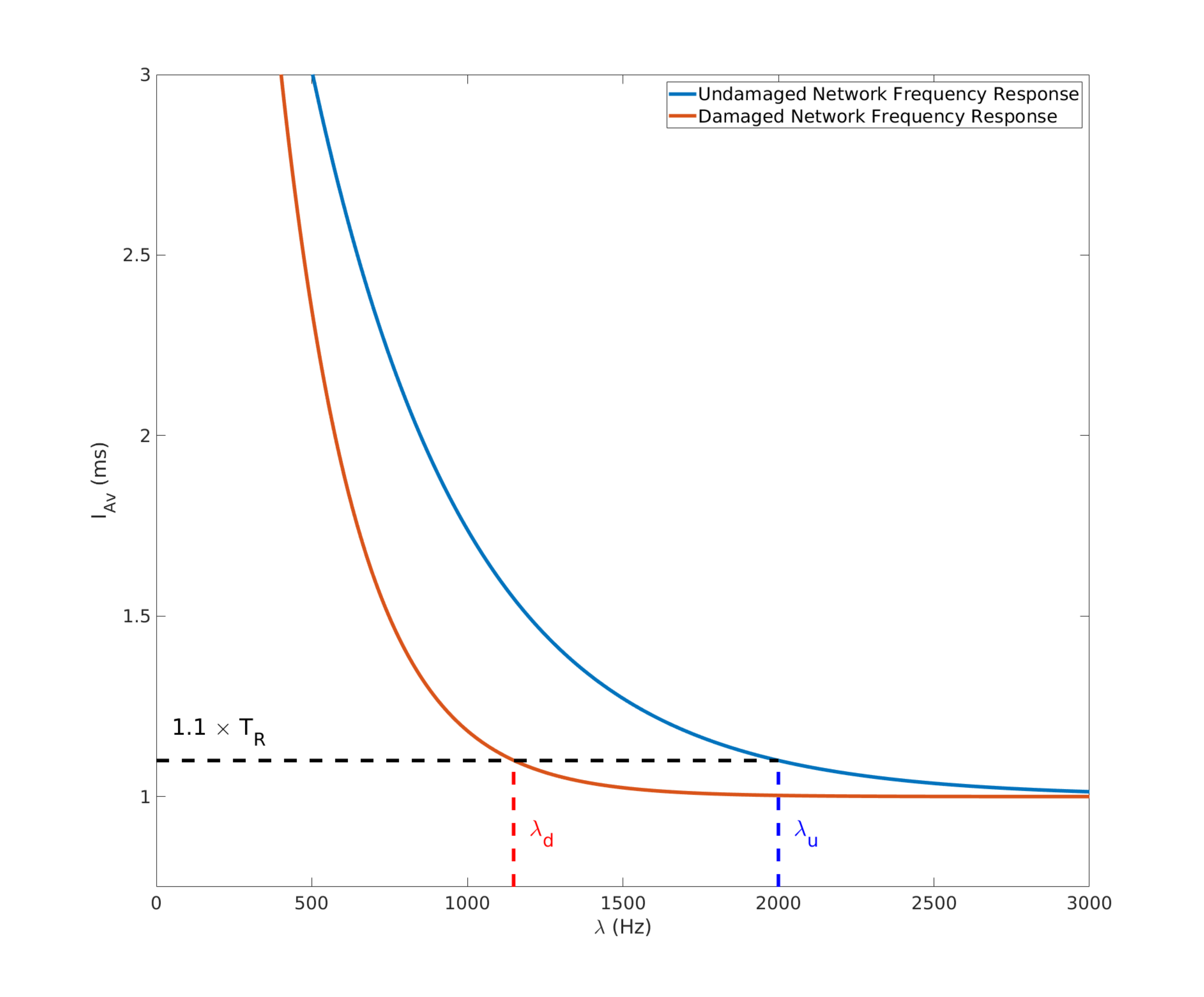}
\caption{Prototypical curves for the average interspike interval as a function of frequency for associated undamaged and damaged networks. They decrease monotonically towards an asymptote at  $T_R$. We also have marked $\lambda_u$ and $\lambda_d$, the cutoff frequencies for these networks where the interspike interval reaches 1.1$\times T_R$. Note that while in this illustration $\lambda_d<\lambda_u$, this is not always the case. Our damage metric $d_{BW}$ measures absolute bandwidth difference, so it does not discriminate between the cases where damage incurs a decrease or increase in bandwidth.}
\label{fig:protocurves}
\end{figure}

{The prototypical frequency response curve is shown in Figure \ref{fig:protocurves}. Such a curve will monotonically decrease towards a horizontal asymptote. Thus, the network's capacity to discriminate between different inputs diminishes as the input frequency increases. In this case, the horizontal asymptote is at $T_R$, which is the theoretical lower bound of the output spiking period.}

\par{We define a \textit{cutoff frequency} above which we say the network poorly discriminates between input stimulus frequencies. Every network comes with an associated cutoff frequency. When a network is damaged, it stands to reason that the cutoff frequency will change. This change in cutoff frequency depends on both the damage paradigm and the original architecture to which damage is applied. We are interested in how network architecture itself, as well as modification to it via addition of connections and cells, result in robustness to the effect of damage in this sense.}

\par{For an undamaged network of some architecture, we define the \textit{undamaged network cutoff frequency} $\lambda_u$ by
\begin{equation}
    \lambda_u = \inf\{\lambda \:\; \lvert \:\; I_{av}(f) \leq (1.1)T_R,  \: \forall \:  f >\lambda\}.
\end{equation}
Analogously, let the \textit{damaged network cutoff frequency} $\lambda_d$ be such a frequency for the associated damaged network of the same architecture under a given damage paradigm. Figure \ref{fig:protocurves} also shows prototypical $I_{av}(\lambda)$ curves for associated undamaged and damaged networks with $\lambda_u$ and $\lambda_d$.}

\par{We define a network's \textit{bandwidth} as the length of the frequency interval over which the network discriminates faithfully. We define the \textit{bandwidth damage metric}, $d_{BW}$, as
\begin{equation}
	d_{BW} =  \bigg\lvert\frac{\lambda_d - \lambda_u}{\lambda_u}\bigg\rvert \times 100.
\end{equation}
This is the absolute percent deviation of the damaged bandwidth from the undamaged bandwidth incurred by the specified damage paradigm. That is, it measures how much damage impacts the ability of a network to discriminate between high-frequency stimuli.}

\subsection{Network architectures of focus}\label{methods::architecture}

\par{We consider networks of neurons composed of two-cell feedforward layers in which the layers are also connected by a feedforward structure. This special case of networks is of relevance, as feedforward layered integrate-and-fire models have been used in the past to approach issues such as logic gating and schizophrenia~\cite{Vogels2005,VogelsRajanAbbott2005,VogelsAbbott2007}. The spiking properties of such networks are observed by measuring the output of the last neuron with respect to the order imposed by the feedforward structure within and between layers. To begin, we observe the undamaged network's response. We then compare this to the response of an associated network featuring damage on all axons within layers, but not those between layers. We record from the \textit{same} neuron to obtain the damaged response.}
\par{We then consider layer-wise modifications to the network, such as the additions of feedback paths or neurons. The undamaged and damaged responses of this modified network are determined as well. In each such instance, damage is applied only to the axons that were damaged in the unmodified network.}

\subsubsection*{Two-cell networks}
\par{The one-layer case is that of the two-cell network. We consider the $d_{BW}$ incurred by damaging the single axon in this network, where the output is observed from the second neuron in the feedforward chain.}

\par
\begin{figure}
\centering
\includegraphics[width=0.5\textwidth]{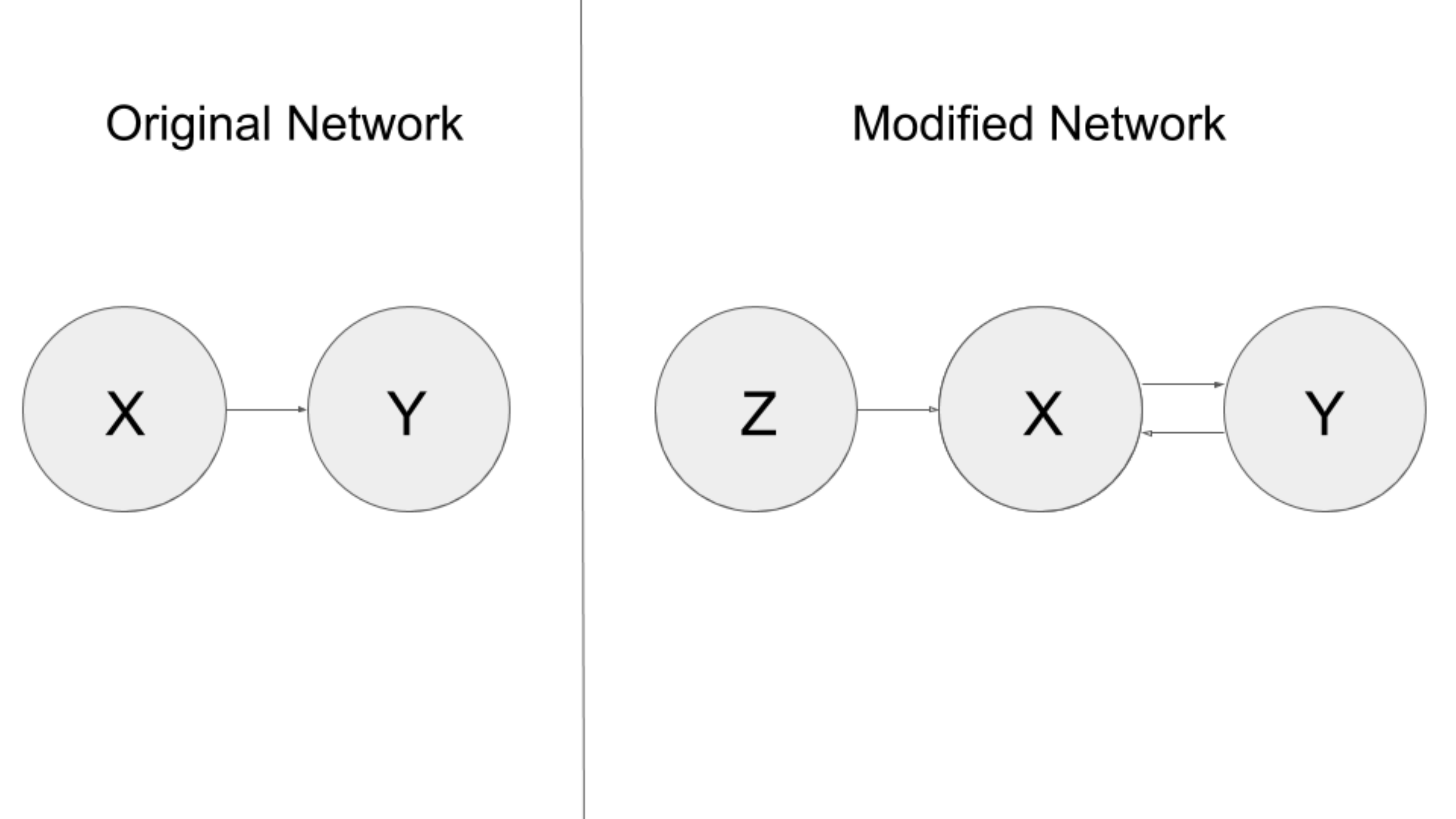}
\caption{The left-hand side shows an arbitrary feedforward network consisting of two cells. $X$ and $Y$ represent the type of cell present at that node (inhibitory or excitatory), and $X$ can be the same as $Y$. The right-hand side shows modifications to the network, where $Z$ is the type of the potentially added neuron. Arrows with white arrowheads represent connections which can be made so as to modify the network -- a feedback path can be added, a third neuron can be added at the front of the chain, or both can be added at once.}
\label{fig:twocellmod}
\end{figure}

{We consider two modifications to networks of this type: (1) the addition of an axon in feedback from the second neuron to the first, and (2) the addition of a neuron at the front of the feedforward chain, connected by a new axon to the first cell in the two-cell feedforward chain. The additional cells in modification (2) can be either inhibitory or excitatory. We also consider the case where both modifications (addition of a feedback path and addition of an additional neuron) take place. As such there are five possible modifications for each two-cell feedforward network. Figure \ref{fig:twocellmod} provides a graphical representation of these modifications made to a given two-cell network. In determining the $d_{BW}$ for these modified networks, we still consider damage only on the single axon which was present in the two-cell network, and we continue to measure the output at the final cell in the feedforward chain.}

\subsubsection*{Layered networks}
\par
\begin{figure}
\centering
\includegraphics[width=0.5\textwidth]{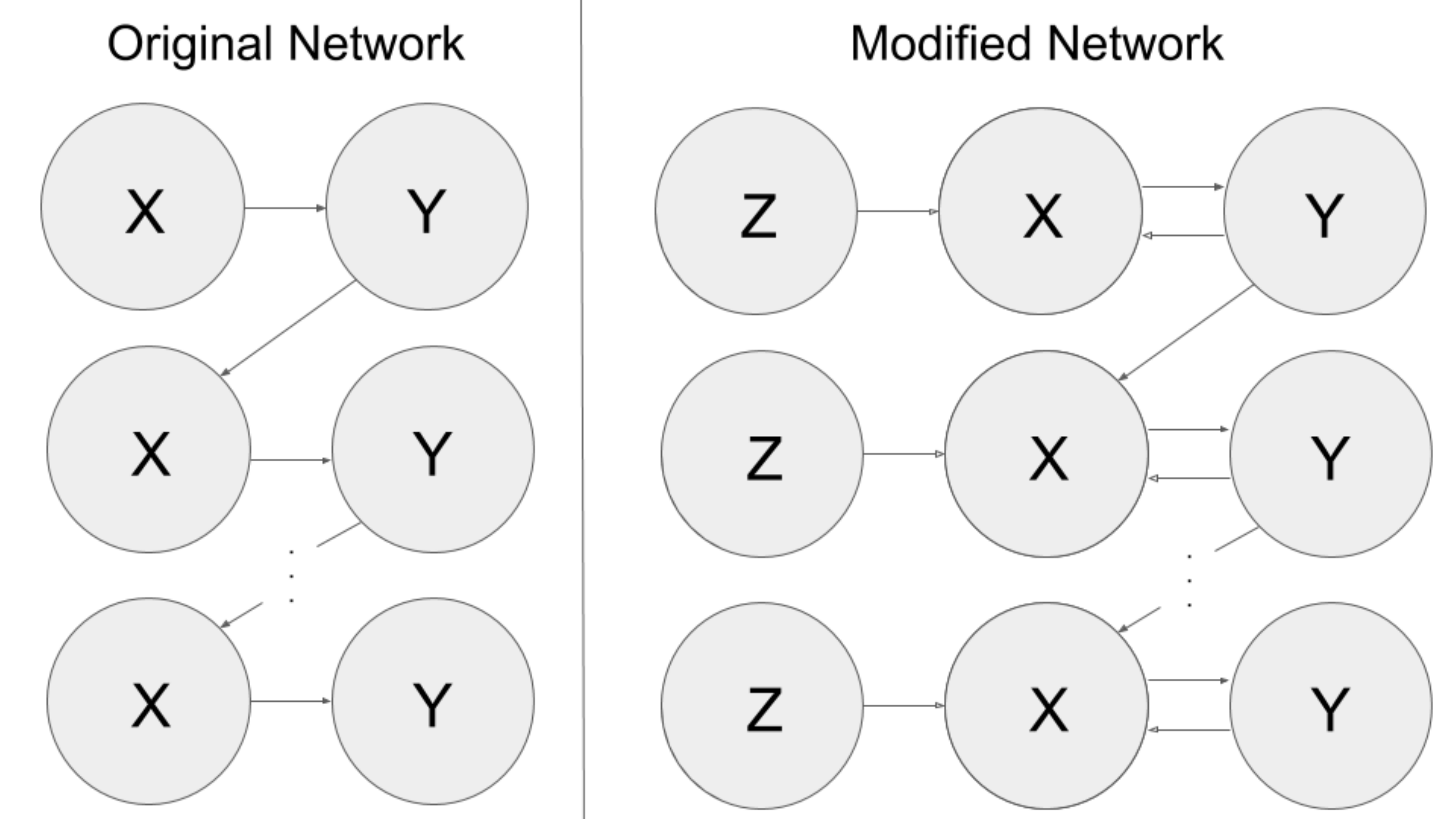}
\caption{The left-hand side shows an arbitrary feedforward network consisting of two-cell layers. The right-hand side shows modifications to the network, which are applied uniformly across layers.}
\label{fig:layeredmod}
\end{figure}

{In studying layered networks, we focus on the \textit{homogeneous} case, in which all layers are identical. The inter-layer connections are realized as an axon connecting the second neuron in a feedforward chain to the first neuron in the next layer's feedforward chain. With $L$ layers, this network is a $2L$-cell feedforward chain, and we read the output from the last cell in this chain. We make modifications of the same 5 types as we did in the two-cell case layer-wise.  Figure \ref{fig:layeredmod} provides a graphical representation of these modifications made to a given layered feedforward network.}

\section{Results and discussion} \label{results}

\par{We present an analysis of the studies described in Section \ref{methods}.  We summarize the results briefly below, and then provide discussion for each network case in detail.}


\par{All investigated networks either benefit from or are unaffected by the addition of inhibitory neurons at the front of each layer with respect to $d_{BW}$-robustness in the face of imposed damage. Due to the exhaustive nature of this study, this provides a plausible, consistent mechanism for the improved robustness of small layered networks, by way of plasticity via the addition of inhibition.}

\par{For single-layer feedforward networks, there always exist modifications which lower the $d_{BW}$. In particular, the addition of an inhibitory cell to the front of a feedforward chain will always result in a smaller $d_{BW}$ than in the base case. Thus, for networks of this type, there is a single mechanism by which neuroplasticity can consistently promote recovery.}

\par{For larger layered networks of the homogeneous type, the addition of an inhibitory cell to the front of each layer will never increase $d_{BW}$ from its base case value. However, it does not necessarily decrease this $d_{BW}$, meaning that this modification will at worst have no effect on the bandwidth of the network.}


\subsection{Single-layer feedforward networks}

\begin{figure}
\centering
    \includegraphics[width=\textwidth]{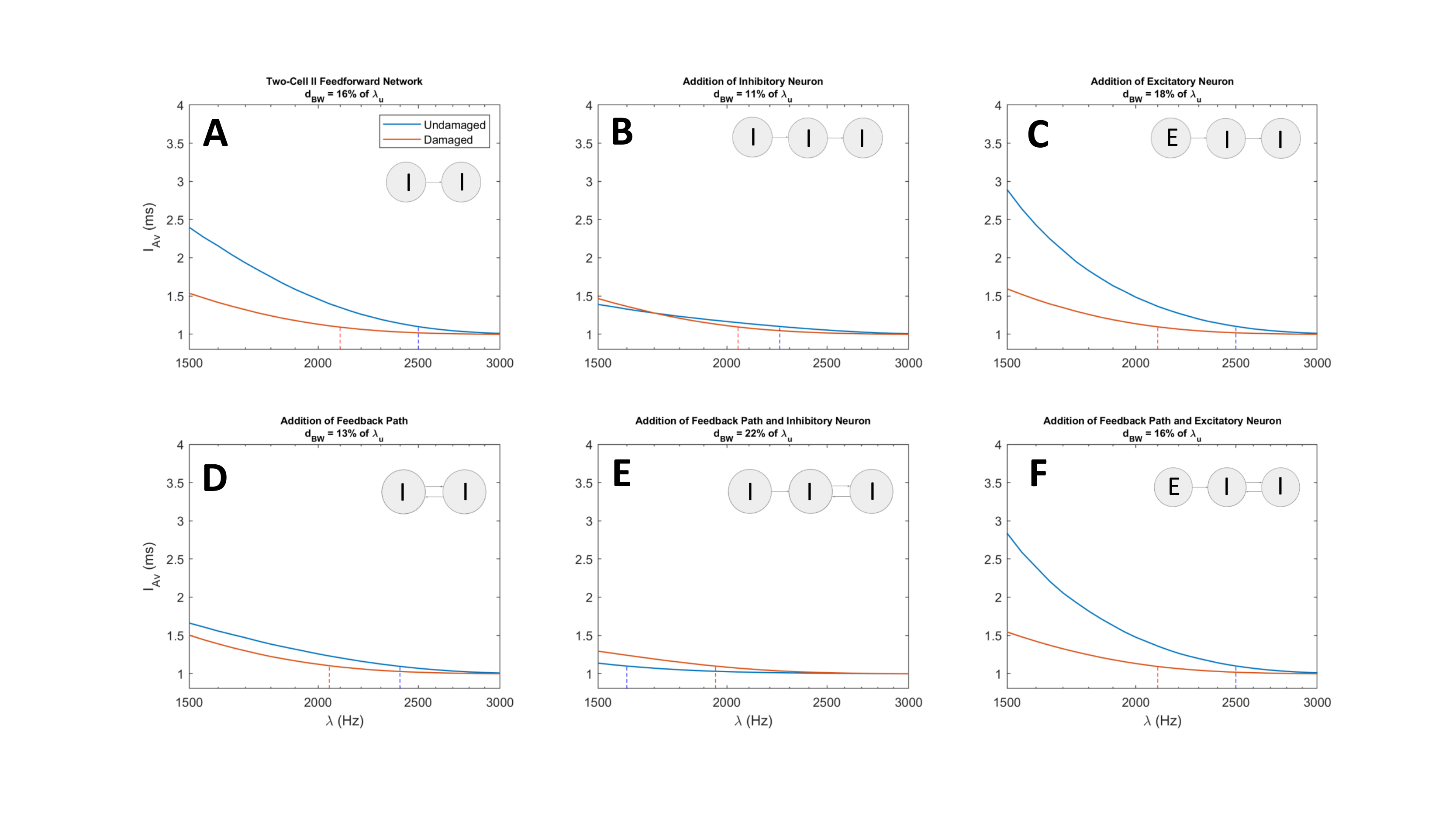}
	\caption{Damaged and undamaged frequency responses of all tested modifications for a single-layer network consisting of two inhibitory neurons. Dotted lines indicates the cutoff frequencies, $\lambda_u$ and $\lambda_d$. \textbf{A} -- The frequency responses for the base case: a two-cell feedforward network. \textbf{B, C} -- The frequency responses for modified networks in which a single neuron is added to the front of the feedforward chain of the base network from that in panel A. It can be seen that addition of an inhibitory cell lowers the $d_{BW}$ value from the base case value, while adding an excitatory cell raises the $d_{BW}$ value. \textbf{D} -- The frequency response of the modified network in which a feedback path is added from the output neuron to the first neuron in the feedforward chain, resulting in a lower $d_{BW}$. \textbf{E, F} -- Frequency responses for modified networks in which both a feedback path and a neuron is added. Addition of feedback and an inhibitory neuron results in a higher $d_{BW}$, whereas addition of feedback and an excitatory neuron incurs no significant change in $d_{BW}$.}
	\label{fig:twotothree}
\end{figure}

\par{In this section we present the exhaustive numerical study of the effects of damage on single-layer feedforward networks. We compare the $d_{BW}$ values for feedforward networks of two cells with corresponding modified networks of two to three cells.}

\par{For example, Figure \ref{fig:twotothree} shows the $\lambda$ vs. $I_{AV}$ curves from which $d_{BW}$ is computed for a feedforward network of two inhibitory cells, as well as the effects of the 5 modifications to such a network.}

\begin{table}
\tiny
\centering
\begin{tabular}{|c|c|c|c|c|c|c|}
\hline
Network & Base Case & Add I Cell & Add E Cell & Add FB & Add FB and I & Add FB and E\\
\hline \hline
EE & 10 & \cellcolor{Gray!25}8 & 24 & 50 & 29 & 67\\
\hline
EI & 10 & 8 \cellcolor{Gray!25}& 24 & 11 & 10 & 0 \cellcolor{Gray!25}\\
\hline
IE & 16 & 11 \cellcolor{Gray!25}& 18 & 18 & 16 & 18\\
\hline
II & 16 & 11 \cellcolor{Gray!25}& 18 & \cellcolor{Gray!25}13 & 22 & 16\\
\hline
\end{tabular}
\caption{$d_{BW}$ values for one-layer networks presented in $\%$ of $\lambda_u$. Highlighted cells show $d_{BW}$ values for modified networks which are lower than the $d_{BW}$ for the respective base case. The addition of an inhibitory cell invariably lowers the $d_{BW}$ from the base case. Similarly, the addition of an excitatory cell in feedforward always raises the $d_{BW}$ from the base case. For an ``EI'' network, the addition of a feedback path as well as an excitatory neuron in feedforward yields a $d_{BW}$ of 0, meaning that for such a modified network, the damaged and undamaged frequency responses are indistinguishable. Addition of a feedback path alone is restorative only for a network consisting of two inhibitory neurons.}
\label{table:layer1}
\end{table}

\par{Table \ref{table:layer1} contains all such $d_{BW}$ comparisons for the single-layer case. Note that the addition of an inhibitory cell always has a restorative effect on the network's $d_{BW}$ -- that is, $d_{BW}$ is always decreased by addition of an inhibitory cell. From an intuitive perspective, the addition of an inhibitory cell at the front of a network decreases the firing rate of the cell to which it is connected. We expect the damage to act as a non-linear low-pass filter, so a decrease in firing rate would make spikes more likely to be faithfully transmitted through a damaged axon. Thus, the restorative nature of the addition of an inhibitory cell in feedforward is consistent with intuition.}

\par{In certain networks, other modifications outperform the addition of an inhibitory cell. For example, in the case of EI feedforward networks, the addition of both a feedback connection and an excitatory neuron entirely eliminates changes in the bandwidth relative to the precision of our simulations. The restorative mechanism of such a modification is not particularly intuitive as a result of the competing effects that are present. That is, excitation at the front of a feedforward network should increase the firing rate of the first neuron and make damage more impactful, but the feedback acts so as to lower the firing rate in a manner proportional to the firing rate itself. In this case, these competing effects play out in the favor of the network's recovery. It is clear from this example that for networks with any deviation from the two-cell feedforward structure, it is difficult to intuit which modifications may result in a lowering of $d_{BW}$. This modification in particular does not have restorative effects for any other two-cell network architecture. No modification other than the addition of a single inhibitory cell is seen to have a consistent restorative effect on networks of this type.}

\subsection{Homogeneous Layered Feedforward Networks}

\begin{table}
\tiny
\centering
\begin{tabular}{|c|c|c|c|c|c|c|}
\hline
Network & Base Case & Add I Cell & Add E Cell & Add FB & Add FB and I & Add FB and E\\
\hline \hline
EE & 33 & 5 \cellcolor{Gray!25} & 62 & 82 & 24 \cellcolor{Gray!25}  & 67\\
\hline
EI & 0 & 0 & 0 & 0 & 0 & 5\\
\hline
IE & 18 & 16 \cellcolor{Gray!25} & 18 & 18 & 16 \cellcolor{Gray!25}& 18\\
\hline
II & 15 & 0 \cellcolor{Gray!25} & 16 & 0 \cellcolor{Gray!25}& 0 \cellcolor{Gray!25}& 16\\
\hline
\end{tabular}
\caption{$d_{BW}$ values for 2-layer networks presented in $\%$ of $\lambda_u$. Highlighted cells show $d_{BW}$ values for modified networks which are lower than the $d_{BW}$ for the respective base case. The addition of an inhibitory cell at the front of each layer can be seen to provide a restorative effect in three of four networks. In the ``EI'' network, we see that the base $d_{BW}$ is 0. In this case, the addition of an inhibitory cell does not raise the $d_{BW}$. Thus, the layer-wise addition of an inhibitory cell in feedforward has, at worst, a neutral effect on two-layer feedforward networks. Unlike in the single layer case, we see here that there is never a better modification than the addition of inhibitory cells. We also see that the addition of both feedback paths and inhibitory neurons in each layer provides a restorative or neutral effect in all four networks.}
\label{table:layer2}
\end{table}

\subsubsection*{Two-layer networks}

\par{In this section we present the effects of damage on layered networks of the homogeneous type. To begin, we consider two-layered networks, for which Table \ref{table:layer2} shows the results of all possible modifications.}

\par{Note that there are networks which show an increase in $d_{BW}$ in response to modifications that have restorative effects in their respective two-cell analogues. For example, we saw that the addition of a feedback path and an excitatory cell had restorative effects on the EI feedforward network of the two-cell type, yet the analogous modification to the two-layer network creates the largest observed increase to $d_{BW}$ across all modifications. This indicates that layered networks do not inherit the characteristics of their single-layer analogues.}

\par{The addition of an inhibitory cell to each layer of a two-layered homogeneous network never increases $d_{BW}$, but it also does not always decrease it. This is weaker than the result seen for two-cell feedforward networks, but still allows for a mechanism by which neuroplasticity can attempt to restore a network of this type without risk of detrimental effects. This provides an instance in which two-layer networks inherit characteristics from their one-layer analogues, despite this not being the case in general.}

\par{We note also that there are two-layer networks which are very robust to damage with respect to this metric. For example, the network which is composed of EI layers experiences no change in bandwidth due to the addition of damage.}

\begin{table}
\tiny
\centering
\begin{tabular}{|c|c|c|c|c|c|c|}
\hline
Network & Base Case & Add I Cell & Add E Cell & Add FB & Add FB and I & Add FB and E\\
\hline \hline
EE & 43 & 24 \cellcolor{Gray!25}& 67 & 82 & 67 & 82\\
\hline
EI & 0 & 0 & 5 & 0 & 0 & 0\\
\hline
IE & 18 & 16 \cellcolor{Gray!25}& 18 & 18 & 18 & 18\\
\hline
II & 15 & 0 \cellcolor{Gray!25}& 18 & 0 \cellcolor{Gray!25}& 0 \cellcolor{Gray!25}& 18\\
\hline
\end{tabular}
\caption{$d_{BW}$ values for 10-layer networks presented in $\%$ of $\lambda_u$. Highlighted cells show $d_{BW}$ values for modified networks which are lower than the $d_{BW}$ for the respective base case. Layer-wise addition of inhibitory neurons has a restorative effect in three of four networks, only having a neutral effect on a network with a base $d_{BW}$ of 0. Furthermore, in all cases there are no modifications which provide a more restorative effect than the addition of inhibitory cells layer-wise.}
\label{table:layer10}
\end{table}

\subsubsection*{Networks containing more layers}
\par{We have recognized a pattern in which the layer-wise addition of inhibitory drive enhances a network's robustness to damage in the one- and two-layer cases. It is natural to ask whether this pattern will continue as the number of layers is increased. Since the complexity of networks increases with the number of layers, we illustrate the potential effects of modification for larger layered networks by focusing on case studies of ten and fifty layers.}
\par{Table  \ref{table:layer10} shows the $d_{BW}$ for ten-layer networks. At this size, it is still the case that layer-wise addition of inhibitory drive never increases a network's $d_{BW}$. Fifty-layer networks exhibit this property as well, as seen in Table \ref{table:layer50}. That is, for \textit{all} tested sizes of homogeneous feedforward networks, we observe a consistent restorative effect of additional layer-wise inhibitory drive.}

\begin{table}
\tiny
\centering
\begin{tabular}{|c|c|c|}
\hline
Network & Base Case & Add I Cell \\
\hline \hline
EE & 43 & 24\\
\hline
EI & 0 & 0\\
\hline
IE & 18 & 16\\
\hline
II & 0 & 0\\
\hline
\end{tabular}
\caption{$d_{BW}$ Values for 50-layer networks in which only the addition of an inhibitory cell modification is made, presented in $\%$ of $\lambda_u$. These networks see at worst neutral effects from the addition of inhibitory neurons layer-wise, where this modification is only neutral when the base network has a $d_{BW}$ of 0. Two of four base networks show $d_{BW}$ values of 0, while in the two networks which present non-zero $d_{BW}$, we observe $d_{BW}$ values of a similar magnitude to those observed in smaller networks.}
\label{table:layer50}
\end{table}






\section{Conclusions} \label{conclusions}
\par{For homogeneous layered networks of varying sizes, we find that there is always a modification that can diminish the effect of damage as captured by the $d_{BW}$ metric. Biophysically, these modifications can be interpreted as the rerouting of both extra- and intra-network connections via neuroplasticity. Thus, neuroplasticity may always provide a mechanism for the mitigation of the effects of damage incurred by focal axonal swelling.}
\par{In particular, we see that there exists a single modification -- the layer-wise addition of inhibition -- that always has a restorative effect on networks of this type. As this modification's effect is present across networks of all tested sizes, it provides a simple mechanism through which neuroplasticity may aid any homogeneous feedforward network.
Other restorative modifications are present in specific network architectures and at specific network sizes. For example, in the single-layer ``EI'' case,  the addition of a feedback connection and an excitatory drive can reduce $d_{BW}$ to 0. Meanwhile, this modification exacerbates damage in all other networks of this size, and in all tested larger networks. Similarly, while the addition of feedback and inhibition has at worst neutral effects on two-layer networks, it can exacerbate damage in networks of other sizes. Thus, the $d_{BW}$ metric is sensitive to perturbations of both network size and network architecture. Besides layer-wise addition of inhibition, there is no tested modification that acts in a restorative fashion on networks of all architectures and sizes. }
\par{Network size, in the sense of number of layers, also plays a restorative role for certain architectures. For example, the ``EI'' network presents a base $d_{BW}$ of 0 at two layers, ten layers and fifty layers. However, network size does not have a restorative effect in general, as ``EE'' and ``IE'' networks present base $d_{BW}$ values that never decrease as network size is increased.}
\par{Our results show mechanisms by which networks of a specific architecture can recover from focal axonal swelling. It is natural to consider similar modifications made to networks of similar architectures. For example, one could consider layered networks that do not have a feedforward structure, or heterogeneous layered networks.}
\par{Our results on homogeneous networks imply that architecture plays a critical role in the base resilience of a network to damage. Naturally, one could consider the set of all small perturbations to a given network and study their effects systematically and exhaustively. Such a study could potentially isolate aspects of architectures which make networks more or less susceptible to damage. With the use of DAPA, such a study is computationally tractable even for large networks -- even ones that feature more complicated neuron dynamics models at the cell level.}

\bibliography{Frost-Mintchev2022.bib}


\begin{thebibliography}{37}
\ifx \bisbn   \undefined \def \bisbn  #1{ISBN #1}\fi
\ifx \binits  \undefined \def \binits#1{#1}\fi
\ifx \bauthor  \undefined \def \bauthor#1{#1}\fi
\ifx \batitle  \undefined \def \batitle#1{#1}\fi
\ifx \bjtitle  \undefined \def \bjtitle#1{#1}\fi
\ifx \bvolume  \undefined \def \bvolume#1{\textbf{#1}}\fi
\ifx \byear  \undefined \def \byear#1{#1}\fi
\ifx \bissue  \undefined \def \bissue#1{#1}\fi
\ifx \bfpage  \undefined \def \bfpage#1{#1}\fi
\ifx \blpage  \undefined \def \blpage #1{#1}\fi
\ifx \burl  \undefined \def \burl#1{\textsf{#1}}\fi
\ifx \doiurl  \undefined \def \doiurl#1{\url{https://doi.org/#1}}\fi
\ifx \betal  \undefined \def \betal{\textit{et al.}}\fi
\ifx \binstitute  \undefined \def \binstitute#1{#1}\fi
\ifx \binstitutionaled  \undefined \def \binstitutionaled#1{#1}\fi
\ifx \bctitle  \undefined \def \bctitle#1{#1}\fi
\ifx \beditor  \undefined \def \beditor#1{#1}\fi
\ifx \bpublisher  \undefined \def \bpublisher#1{#1}\fi
\ifx \bbtitle  \undefined \def \bbtitle#1{#1}\fi
\ifx \bedition  \undefined \def \bedition#1{#1}\fi
\ifx \bseriesno  \undefined \def \bseriesno#1{#1}\fi
\ifx \blocation  \undefined \def \blocation#1{#1}\fi
\ifx \bsertitle  \undefined \def \bsertitle#1{#1}\fi
\ifx \bsnm \undefined \def \bsnm#1{#1}\fi
\ifx \bsuffix \undefined \def \bsuffix#1{#1}\fi
\ifx \bparticle \undefined \def \bparticle#1{#1}\fi
\ifx \barticle \undefined \def \barticle#1{#1}\fi
\bibcommenthead
\ifx \bconfdate \undefined \def \bconfdate #1{#1}\fi
\ifx \botherref \undefined \def \botherref #1{#1}\fi
\ifx \url \undefined \def \url#1{\textsf{#1}}\fi
\ifx \bchapter \undefined \def \bchapter#1{#1}\fi
\ifx \bbook \undefined \def \bbook#1{#1}\fi
\ifx \bcomment \undefined \def \bcomment#1{#1}\fi
\ifx \oauthor \undefined \def \oauthor#1{#1}\fi
\ifx \citeauthoryear \undefined \def \citeauthoryear#1{#1}\fi
\ifx \endbibitem  \undefined \def \endbibitem {}\fi
\ifx \bconflocation  \undefined \def \bconflocation#1{#1}\fi
\ifx \arxivurl  \undefined \def \arxivurl#1{\textsf{#1}}\fi
\csname PreBibitemsHook\endcsname

\bibitem{Maia2014}
\begin{barticle}
\bauthor{\bsnm{Maia}, \binits{P.D.}},
\bauthor{\bsnm{Kutz}, \binits{J.N.}}:
\batitle{Compromised axonal functionality after neurodegeneration, concussion
  and/or traumatic brain injury}.
\bjtitle{Journal of Computational Neuroscience}
\bvolume{37}(\bissue{2}),
\bfpage{317}--\blpage{332}
(\byear{2014}).
\doiurl{10.1007/s10827-014-0504-x}
\end{barticle}
\endbibitem

\bibitem{Johnson2013}
\begin{barticle}
\bauthor{\bsnm{Johnson}, \binits{V.E.}},
\bauthor{\bsnm{Stewart}, \binits{W.}},
\bauthor{\bsnm{Smith}, \binits{D.H.}}:
\batitle{Axonal pathology in traumatic brain injury}.
\bjtitle{Experimental Neurology}
\bvolume{246},
\bfpage{35}--\blpage{43}
(\byear{2013}).
\doiurl{10.1016/j.expneurol.2012.01.013}
\end{barticle}
\endbibitem

\bibitem{Tagge2018}
\begin{barticle}
\bauthor{\bsnm{Tagge}, \binits{C.A.}},
\bauthor{\bsnm{Fisher}, \binits{A.M.}},
\bauthor{\bsnm{Minaeva}, \binits{O.V.}},
\bauthor{\bsnm{Gaudreau-Balderrama}, \binits{A.}},
\bauthor{\bsnm{Moncaster}, \binits{J.A.}},
\bauthor{\bsnm{Zhang}, \binits{X.-L.}},
\bauthor{\bsnm{Wojnarowicz}, \binits{M.W.}},
\bauthor{\bsnm{Casey}, \binits{N.}},
\bauthor{\bsnm{Lu}, \binits{H.}},
\bauthor{\bsnm{Kokiko-Cochran}, \binits{O.N.}},
\bauthor{\bsnm{Saman}, \binits{S.}},
\bauthor{\bsnm{Ericsson}, \binits{M.}},
\bauthor{\bsnm{Onos}, \binits{K.D.}},
\bauthor{\bsnm{Veksler}, \binits{R.}},
\bauthor{\bsnm{Senatorov}, \binits{V.V.}},
\bauthor{\bsnm{Kondo}, \binits{A.}},
\bauthor{\bsnm{Zhou}, \binits{X.Z.}},
\bauthor{\bsnm{Miry}, \binits{O.}},
\bauthor{\bsnm{Vose}, \binits{L.R.}},
\bauthor{\bsnm{Gopaul}, \binits{K.R.}},
\bauthor{\bsnm{Upreti}, \binits{C.}},
\bauthor{\bsnm{Nowinski}, \binits{C.J.}},
\bauthor{\bsnm{Cantu}, \binits{R.C.}},
\bauthor{\bsnm{Alvarez}, \binits{V.E.}},
\bauthor{\bsnm{Hildebrandt}, \binits{A.M.}},
\bauthor{\bsnm{Franz}, \binits{E.S.}},
\bauthor{\bsnm{Konrad}, \binits{J.}},
\bauthor{\bsnm{Hamilton}, \binits{J.A.}},
\bauthor{\bsnm{Hua}, \binits{N.}},
\bauthor{\bsnm{Tripodis}, \binits{Y.}},
\bauthor{\bsnm{Anderson}, \binits{A.T.}},
\bauthor{\bsnm{Howell}, \binits{G.R.}},
\bauthor{\bsnm{Kaufer}, \binits{D.}},
\bauthor{\bsnm{Hall}, \binits{G.F.}},
\bauthor{\bsnm{Lu}, \binits{K.P.}},
\bauthor{\bsnm{Ransohoff}, \binits{R.M.}},
\bauthor{\bsnm{Cleveland}, \binits{R.O.}},
\bauthor{\bsnm{Kowall}, \binits{N.W.}},
\bauthor{\bsnm{Stein}, \binits{T.D.}},
\bauthor{\bsnm{Lamb}, \binits{B.T.}},
\bauthor{\bsnm{Huber}, \binits{B.R.}},
\bauthor{\bsnm{Moss}, \binits{W.C.}},
\bauthor{\bsnm{Friedman}, \binits{A.}},
\bauthor{\bsnm{Stanton}, \binits{P.K.}},
\bauthor{\bsnm{McKee}, \binits{A.C.}},
\bauthor{\bsnm{Goldstein}, \binits{L.E.}}:
\batitle{Concussion, microvascular injury, and early tauopathy in young
  athletes after impact head injury and an impact concussion mouse model}.
\bjtitle{Brain}
\bvolume{141}(\bissue{2}),
\bfpage{422}--\blpage{458}
(\byear{2018}).
\doiurl{10.1093/brain/awx350}
\end{barticle}
\endbibitem

\bibitem{Maxwell1997}
\begin{barticle}
\bauthor{\bsnm{Maxwell}, \binits{W.L.}},
\bauthor{\bsnm{Povlishock}, \binits{J.T.}},
\bauthor{\bsnm{Graham}, \binits{D.L.}}:
\batitle{A mechanistic analysis of nondisruptive axonal injury: A review}.
\bjtitle{Journal of Neurotrauma}
\bvolume{14}(\bissue{7}),
\bfpage{419}--\blpage{440}
(\byear{1997}).
\doiurl{10.1089/neu.1997.14.419}
\end{barticle}
\endbibitem

\bibitem{Wu2021}
\begin{barticle}
\bauthor{\bsnm{Wu}, \binits{Y.-H.}},
\bauthor{\bsnm{Rosset}, \binits{S.}},
\bauthor{\bsnm{Lee}, \binits{T.-r.}},
\bauthor{\bsnm{Dragunow}, \binits{M.}},
\bauthor{\bsnm{Park}, \binits{T.}},
\bauthor{\bsnm{Shim}, \binits{V.}}:
\batitle{\emph{In vitro} models of traumatic brain injury: A systematic
  review}.
\bjtitle{Journal of Neurotrauma}
\bvolume{38}(\bissue{17}),
\bfpage{2336}--\blpage{2372}
(\byear{2021}).
\doiurl{10.1089/neu.2020.7402}
\end{barticle}
\endbibitem

\bibitem{Tang-Schomer2012}
\begin{barticle}
\bauthor{\bsnm{Tang-Schomer}, \binits{M.D.}},
\bauthor{\bsnm{Johnson}, \binits{V.E.}},
\bauthor{\bsnm{Baas}, \binits{P.W.}},
\bauthor{\bsnm{Stewart}, \binits{W.}},
\bauthor{\bsnm{Smith}, \binits{D.H.}}:
\batitle{Partial interruption of axonal transport due to microtubule breakage
  accounts for the formation of periodic varicosities after traumatic axonal
  injury}.
\bjtitle{Experimental Neurology}
\bvolume{233}(\bissue{1}),
\bfpage{364}--\blpage{372}
(\byear{2012}).
\doiurl{10.1016/j.expneurol.2011.10.030}
\end{barticle}
\endbibitem

\bibitem{Wang2011}
\begin{barticle}
\bauthor{\bsnm{Wang}, \binits{J.}},
\bauthor{\bsnm{Hamm}, \binits{R.J.}},
\bauthor{\bsnm{Povlishock}, \binits{J.T.}}:
\batitle{Traumatic axonal injury in the optic nerve: Evidence for axonal
  swelling, disconnection, dieback, and reorganization}.
\bjtitle{Journal of Neurotrauma}
\bvolume{28}(\bissue{7}),
\bfpage{1185}--\blpage{1198}
(\byear{2011}).
\doiurl{10.1089/neu.2011.1756}
\end{barticle}
\endbibitem

\bibitem{Maia2013}
\begin{barticle}
\bauthor{\bsnm{Maia}, \binits{P.D.}},
\bauthor{\bsnm{Kutz}, \binits{J.N.}}:
\batitle{Identifying critical regions for spike propagation in axon segments}.
\bjtitle{Journal of Computational Neuroscience}
\bvolume{36}(\bissue{2}),
\bfpage{141}--\blpage{155}
(\byear{2013}).
\doiurl{10.1007/s10827-013-0459-3}
\end{barticle}
\endbibitem

\bibitem{Debanne2011}
\begin{barticle}
\bauthor{\bsnm{Debanne}, \binits{D.}},
\bauthor{\bsnm{Campanac}, \binits{E.}},
\bauthor{\bsnm{Bialowas}, \binits{A.}},
\bauthor{\bsnm{Carlier}, \binits{E.}},
\bauthor{\bsnm{Alcaraz}, \binits{G.}}:
\batitle{Axon physiology}.
\bjtitle{Physiological Reviews}
\bvolume{91}(\bissue{2}),
\bfpage{555}--\blpage{602}
(\byear{2011}).
\doiurl{10.1152/physrev.00048.2009}
\end{barticle}
\endbibitem

\bibitem{Maia2015}
\begin{barticle}
\bauthor{\bsnm{Maia}, \binits{P.D.}},
\bauthor{\bsnm{Hemphill}, \binits{M.A.}},
\bauthor{\bsnm{Zehnder}, \binits{B.}},
\bauthor{\bsnm{Zhang}, \binits{C.}},
\bauthor{\bsnm{Parker}, \binits{K.K.}},
\bauthor{\bsnm{Kutz}, \binits{J.N.}}:
\batitle{Diagnostic tools for evaluating the impact of focal axonal swellings
  arising in neurodegenerative diseases and/or traumatic brain injury}.
\bjtitle{Journal of Neuroscience Methods}
\bvolume{253},
\bfpage{233}--\blpage{243}
(\byear{2015}).
\doiurl{10.1016/j.jneumeth.2015.06.022}
\end{barticle}
\endbibitem

\bibitem{Manor1991}
\begin{barticle}
\bauthor{\bsnm{Manor}, \binits{Y.}},
\bauthor{\bsnm{Koch}, \binits{C.}},
\bauthor{\bsnm{Segev}, \binits{I.}}:
\batitle{Effect of geometrical irregularities on propagation delay in axonal
  trees}.
\bjtitle{Biophysical Journal}
\bvolume{60}(\bissue{6}),
\bfpage{1424}--\blpage{1437}
(\byear{1991}).
\doiurl{10.1016/s0006-3495(91)82179-8}
\end{barticle}
\endbibitem

\bibitem{Ramon1975}
\begin{botherref}
\oauthor{\bsnm{Ram{\'{o}}n}, \binits{F.}},
\oauthor{\bsnm{Joyner}, \binits{R.W.}},
\oauthor{\bsnm{Moore}, \binits{J.W.}}:
Propagation of action potentials in inhomogeneous axon regions,
pp. 85--100.
Springer
(1975).
\doiurl{10.1007/978-1-4684-2637-3_8}.
\url{https://doi.org/10.1007%2F978-1-4684-2637-3_8}
\end{botherref}
\endbibitem

\bibitem{Maia2017}
\begin{barticle}
\bauthor{\bsnm{Maia}, \binits{P.D.}},
\bauthor{\bsnm{Kutz}, \binits{J.N.}}:
\batitle{Reaction time impairments in decision-making networks as a diagnostic
  marker for traumatic brain injuries and neurological diseases}.
\bjtitle{Journal of Computational Neuroscience}
\bvolume{42}(\bissue{3}),
\bfpage{323}--\blpage{347}
(\byear{2017}).
\doiurl{10.1007/s10827-017-0643-y}
\end{barticle}
\endbibitem

\bibitem{Sharp2014}
\begin{barticle}
\bauthor{\bsnm{Sharp}, \binits{D.J.}},
\bauthor{\bsnm{Scott}, \binits{G.}},
\bauthor{\bsnm{Leech}, \binits{R.}}:
\batitle{Network dysfunction after traumatic brain injury}.
\bjtitle{Nature Reviews Neurology}
\bvolume{10}(\bissue{3}),
\bfpage{156}--\blpage{166}
(\byear{2014}).
\doiurl{10.1038/nrneurol.2014.15}
\end{barticle}
\endbibitem

\bibitem{Rudy2016}
\begin{barticle}
\bauthor{\bsnm{Rudy}, \binits{S.}},
\bauthor{\bsnm{Maia}, \binits{P.D.}},
\bauthor{\bsnm{Kutz}, \binits{J.N.}}:
\batitle{Cognitive and behavioral deficits arising from neurodegeneration and
  traumatic brain injury: a model for the underlying role of focal axonal
  swellings in neuronal networks with plasticity}.
\bjtitle{Journal of Systems and Integrative Neuroscience}
\bvolume{2}(\bissue{2}),
\bfpage{114}--\blpage{121}
(\byear{2016}).
\doiurl{10.15761/jsin.1000120}
\end{barticle}
\endbibitem

\bibitem{Sussillo2009}
\begin{barticle}
\bauthor{\bsnm{Sussillo}, \binits{D.}},
\bauthor{\bsnm{Abbott}, \binits{L.F.}}:
\batitle{Generating coherent patterns of activity from chaotic neural
  networks}.
\bjtitle{Neuron}
\bvolume{63}(\bissue{4}),
\bfpage{544}--\blpage{557}
(\byear{2009}).
\doiurl{10.1016/j.neuron.2009.07.018}
\end{barticle}
\endbibitem

\bibitem{Lusch2018}
\begin{barticle}
\bauthor{\bsnm{Lusch}, \binits{B.}},
\bauthor{\bsnm{Weholt}, \binits{J.}},
\bauthor{\bsnm{Maia}, \binits{P.D.}},
\bauthor{\bsnm{Kutz}, \binits{J.N.}}:
\batitle{Modeling cognitive deficits following neurodegenerative diseases and
  traumatic brain injuries with deep convolutional neural networks}.
\bjtitle{Brain and Cognition}
\bvolume{123},
\bfpage{154}--\blpage{164}
(\byear{2018}).
\doiurl{10.1016/j.bandc.2018.02.012}
\end{barticle}
\endbibitem

\bibitem{Ermentrout2010}
\begin{bbook}
\bauthor{\bsnm{Ermentrout}, \binits{G.B.}},
\bauthor{\bsnm{Terman}, \binits{D.H.}}:
\bbtitle{Mathematical Foundations of Neuroscience}.
\bpublisher{Springer}, \blocation{???}
(\byear{2010}).
\doiurl{10.1007/978-0-387-87708-2}.
\burl{https://doi.org/10.1007
\end{bbook}
\endbibitem

\bibitem{Herculano_Houzel2009}
\begin{botherref}
\oauthor{\bsnm{Herculano-Houzel}, \binits{S.}}:
The human brain in numbers: a linearly scaled-up primate brain.
Frontiers in Human Neuroscience
\textbf{3}
(2009).
\doiurl{10.3389/neuro.09.031.2009}
\end{botherref}
\endbibitem

\bibitem{Ger-2002}
\begin{bbook}
\bauthor{\bsnm{Gerstner}, \binits{W.}},
\bauthor{\bsnm{Kistler}, \binits{W.M.}}:
\bbtitle{Spiking Neuron Models: Single Neurons, Populations, Plasticity}.
\bpublisher{Cambridge University Press},
\blocation{Cambridge, U.K. New York}
(\byear{2002}).
\burl{https://books.google.com/books?id=Rs4oc7HfxIUC}
\end{bbook}
\endbibitem

\bibitem{Izhikevich2004}
\begin{barticle}
\bauthor{\bsnm{Izhikevich}, \binits{E.M.}}:
\batitle{Which model to use for cortical spiking neurons?}
\bjtitle{{IEEE} Transactions on Neural Networks}
\bvolume{15}(\bissue{5}),
\bfpage{1063}--\blpage{1070}
(\byear{2004}).
\doiurl{10.1109/tnn.2004.832719}
\end{barticle}
\endbibitem

\bibitem{Vogels2005}
\begin{barticle}
\bauthor{\bsnm{Vogels}, \binits{T.P.}}:
\batitle{Signal propagation and logic gating in networks of integrate-and-fire
  neurons}.
\bjtitle{Journal of Neuroscience}
\bvolume{25}(\bissue{46}),
\bfpage{10786}--\blpage{10795}
(\byear{2005}).
\doiurl{10.1523/jneurosci.3508-05.2005}
\end{barticle}
\endbibitem

\bibitem{VogelsRajanAbbott2005}
\begin{barticle}
\bauthor{\bsnm{Vogels}, \binits{T.P.}},
\bauthor{\bsnm{Rajan}, \binits{K.}},
\bauthor{\bsnm{Abbott}, \binits{L.F.}}, \betal:
\batitle{Neural network dynamics}.
\bjtitle{Annual review of neuroscience}
\bvolume{28},
\bfpage{357}
(\byear{2005})
\end{barticle}
\endbibitem

\bibitem{VogelsAbbott2007}
\begin{barticle}
\bauthor{\bsnm{Vogels}, \binits{T.P.}},
\bauthor{\bsnm{Abbott}, \binits{L.}}:
\batitle{Gating deficits in model networks: a path to schizophrenia?}
\bjtitle{Pharmacopsychiatry}
\bvolume{40}(\bissue{S 1}),
\bfpage{73}--\blpage{77}
(\byear{2007})
\end{barticle}
\endbibitem

\bibitem{Abeles1994}
\begin{botherref}
\oauthor{\bsnm{Abeles}, \binits{M.}}:
Firing rates and weil-timed events in the cerebral cortex,
pp. 121--140.
Springer
(1994).
\doiurl{10.1007/978-1-4612-4320-5_3}.
\url{https://doi.org/10.1007%2F978-1-4612-4320-5_3}
\end{botherref}
\endbibitem

\bibitem{Bialek1991}
\begin{barticle}
\bauthor{\bsnm{Bialek}, \binits{W.}},
\bauthor{\bsnm{Rieke}, \binits{F.}},
\bauthor{\bparticle{de~Ruyter~van} \bsnm{Steveninck}, \binits{R.R.}},
\bauthor{\bsnm{Warland}, \binits{D.}}:
\batitle{Reading a neural code}.
\bjtitle{Science}
\bvolume{252}(\bissue{5014}),
\bfpage{1854}--\blpage{1857}
(\byear{1991}).
\doiurl{10.1126/science.2063199}
\end{barticle}
\endbibitem

\bibitem{Haslinger2010}
\begin{barticle}
\bauthor{\bsnm{Haslinger}, \binits{R.}},
\bauthor{\bsnm{Klinkner}, \binits{K.L.}},
\bauthor{\bsnm{Shalizi}, \binits{C.R.}}:
\batitle{The computational structure of spike trains}.
\bjtitle{Neural Computation}
\bvolume{22}(\bissue{1}),
\bfpage{121}--\blpage{157}
(\byear{2010}).
\doiurl{10.1162/neco.2009.12-07-678}
\end{barticle}
\endbibitem

\bibitem{Lestienne1996}
\begin{barticle}
\bauthor{\bsnm{Lestienne}, \binits{R.}}:
\batitle{Determination of the precision of spike timing in the visual cortex of
  anaesthetised cats}.
\bjtitle{Biological Cybernetics}
\bvolume{74}(\bissue{1}),
\bfpage{55}--\blpage{61}
(\byear{1996}).
\doiurl{10.1007/bf00199137}
\end{barticle}
\endbibitem

\bibitem{Young2022}
\begin{barticle}
\bauthor{\bsnm{Young}, \binits{L.-S.}}:
\batitle{The brain is a dynamical system}.
\bjtitle{SIAM News}
\bvolume{55}(\bissue{6}),
\bfpage{1}--\blpage{2}
(\byear{2022})
\end{barticle}
\endbibitem

\bibitem{Brette2007}
\begin{barticle}
\bauthor{\bsnm{Brette}, \binits{R.}},
\bauthor{\bsnm{Rudolph}, \binits{M.}},
\bauthor{\bsnm{Carnevale}, \binits{T.}},
\bauthor{\bsnm{Hines}, \binits{M.}},
\bauthor{\bsnm{Beeman}, \binits{D.}},
\bauthor{\bsnm{Bower}, \binits{J.M.}},
\bauthor{\bsnm{Diesmann}, \binits{M.}},
\bauthor{\bsnm{Morrison}, \binits{A.}},
\bauthor{\bsnm{Goodman}, \binits{P.H.}},
\bauthor{\bsnm{Harris}, \binits{F.C.}},
\bauthor{\bsnm{Zirpe}, \binits{M.}},
\bauthor{\bsnm{Natschl\"{a}ger}, \binits{T.}},
\bauthor{\bsnm{Pecevski}, \binits{D.}},
\bauthor{\bsnm{Ermentrout}, \binits{B.}},
\bauthor{\bsnm{Djurfeldt}, \binits{M.}},
\bauthor{\bsnm{Lansner}, \binits{A.}},
\bauthor{\bsnm{Rochel}, \binits{O.}},
\bauthor{\bsnm{Vieville}, \binits{T.}},
\bauthor{\bsnm{Muller}, \binits{E.}},
\bauthor{\bsnm{Davison}, \binits{A.P.}},
\bauthor{\bsnm{Boustani}, \binits{S.E.}},
\bauthor{\bsnm{Destexhe}, \binits{A.}}:
\batitle{Simulation of networks of spiking neurons: A review of tools and
  strategies}.
\bjtitle{Journal of Computational Neuroscience}
\bvolume{23}(\bissue{3}),
\bfpage{349}--\blpage{398}
(\byear{2007}).
\doiurl{10.1007/s10827-007-0038-6}
\end{barticle}
\endbibitem

\bibitem{dayan2001theoretical}
\begin{bbook}
\bauthor{\bsnm{Dayan}, \binits{P.}},
\bauthor{\bsnm{Abbott}, \binits{L.F.}}:
\bbtitle{Theoretical Neuroscience : Computational and Mathematical Modeling of
  Neural Systems}.
\bpublisher{MIT Press},
\blocation{Cambridge, Mass}
(\byear{2001})
\end{bbook}
\endbibitem

\bibitem{Rudolph2006}
\begin{barticle}
\bauthor{\bsnm{Rudolph}, \binits{M.}},
\bauthor{\bsnm{Destexhe}, \binits{A.}}:
\batitle{Analytical integrate-and-fire neuron models with conductance-based
  dynamics for event-driven simulation strategies}.
\bjtitle{Neural Computation}
\bvolume{18}(\bissue{9}),
\bfpage{2146}--\blpage{2210}
(\byear{2006}).
\doiurl{10.1162/neco.2006.18.9.2146}
\end{barticle}
\endbibitem

\bibitem{Hansel1998}
\begin{barticle}
\bauthor{\bsnm{Hansel}, \binits{D.}},
\bauthor{\bsnm{Mato}, \binits{G.}},
\bauthor{\bsnm{Meunier}, \binits{C.}},
\bauthor{\bsnm{Neltner}, \binits{L.}}:
\batitle{On numerical simulations of integrate-and-fire neural networks}.
\bjtitle{Neural Computation}
\bvolume{10}(\bissue{2}),
\bfpage{467}--\blpage{483}
(\byear{1998}).
\doiurl{10.1162/089976698300017845}
\end{barticle}
\endbibitem

\bibitem{Hodgkin1952}
\begin{barticle}
\bauthor{\bsnm{Hodgkin}, \binits{A.L.}},
\bauthor{\bsnm{Huxley}, \binits{A.F.}}:
\batitle{A quantitative description of membrane current and its application to
  conduction and excitation in nerve}.
\bjtitle{The Journal of Physiology}
\bvolume{117}(\bissue{4}),
\bfpage{500}--\blpage{544}
(\byear{1952}).
\doiurl{10.1113/jphysiol.1952.sp004764}
\end{barticle}
\endbibitem

\bibitem{Thapa2014}
\begin{barticle}
\bauthor{\bsnm{Thapa}, \binits{N.}},
\bauthor{\bsnm{Gudejko}, \binits{M.}}:
\batitle{Numerical solution of heat equation by spectral method}.
\bjtitle{Applied Mathematical Sciences}
\bvolume{8},
\bfpage{397}--\blpage{404}
(\byear{2014}).
\doiurl{10.12988/ams.2014.39502}
\end{barticle}
\endbibitem

\bibitem{Fornberg1994}
\begin{barticle}
\bauthor{\bsnm{Fornberg}, \binits{B.}},
\bauthor{\bsnm{Sloan}, \binits{D.M.}}:
\batitle{A review of pseudospectral methods for solving partial differential
  equations}.
\bjtitle{Acta Numerica}
\bvolume{3},
\bfpage{203}--\blpage{267}
(\byear{1994}).
\doiurl{10.1017/s0962492900002440}
\end{barticle}
\endbibitem

\bibitem{heeger2000poisson}
\begin{barticle}
\bauthor{\bsnm{Heeger}, \binits{D.}}, \betal:
\batitle{Poisson model of spike generation}.
\bjtitle{Handout, University of Standford}
\bvolume{5}(\bissue{1-13}),
\bfpage{76}
(\byear{2000})
\end{barticle}
\endbibitem

\end{thebibliography}

\appendix
\section*{Appendix on Axonal Modeling}

\subsection*{Pseudo-Spectral Method for Axon PDE Model}

\subsubsection*{The Spectral Method}
The pseudo-spectral method for nonlinear PDE is a modification of the simpler spectral method. For an illustration of the latter, consider a simplified version of the first PDE
\begin{equation}
\frac{\partial V}{\partial t} = \frac{D}{d(x)}\frac{\partial}{\partial x}\bigg(\frac{d^2(x)}{r_L(x)}\frac{\partial V}{\partial x}\bigg).
\end{equation}
One fixes a discretization of space so that at each value of time, $V$ is represented by an $N$-point vector. 

Suppose $\{\psi_n\}_{n=0}^{N-1}$ constitutes an orthonormal basis such as the Fourier basis on the space of discrete-space N-point functions. One assumes that $V$ can be written as
\[
V(x,t) = \sum_{n=0}^{N-1}c_n(t)\psi_n(x).
\]
Of note, the coefficients $c_n$ in this representation do not depend on the spatial coordinate $x$. Using then the standard Fourier basis vectors for discrete-space functions with finite domain, these coefficients are exactly those determined by the discrete Fourier transform of the function in the variable $x$, efficiently computed via use of the fast Fourier transform (FFT) algorithm. The left-hand side of the equation can thus be written as 
\[
\frac{\partial V}{\partial t} = \sum_{n=0}^{N-1}c_n'(t)\psi_n(x).
\]
Appealing to standard properties of the Fourier transform, $x$-derivatives can be taken in the transform domain via multiplication by $i k_n$, where $i$ is the imaginary unit and $k_n$ is the coefficient in the exponent of the $n^{th}$ Fourier basis function. This operation produces
\[
\frac{\partial V}{\partial x} = \sum_{n=0}^{N-1}i k_n c_n(t)\psi_n(x).
\]
The functions $d$ and $r_L$ depend only on space; under the assumption that they have been discretized in the same way, they can also be represented as superpositions of the same Fourier basis functions. Their Fourier coefficients do not depend on time, and as such the only time-dependence in the right-hand side's total Fourier expansion is a result of the Fourier expansion of $V$ alone. Multiplication in real space (the $x$-variable space rather than Fourier space) is given by convolution of Fourier coefficients in Fourier space. As the Fourier coefficients of $r_L$ and $d$ are real numbers, multiplication by these functions in real space simply yields coefficients of the right-hand side of the equation which are linear combinations of the original $c_n(t)$ coefficients. That is to say that the entire PDE can be cast as:
\[
\sum_{n=0}^{N-1}c_n'(t)\psi_n(x) = \sum_{n=0}^{N-1}\bigg(\sum_{m=0}^{N-1}a_{n,m}c_m(t)\bigg)\psi_n(x),
\]
where the $a_{n,m}$ coefficients are complex numbers that do not depend on $t$ or $x$. The $a$ coefficients are determined by: 
\begin{enumerate}
\item Taking the initial FFT of $V$
\item Multiplying by $ik_n$ to obtain the first space derivative of $V$ in the transform domain
\item Taking the inverse FFT (IFFT) and multiplying pointwise by $d^2(x)/r_L(x)$
\item Taking the FFT of this expression, multiplying once again by $ik_n$ to take another derivative
\item Taking another IFFT and multiplying pointwise by $D/d(x)$
\item Taking the FFT of this final expression in real space to obtain the entire right-hand side in Fourier space
\end{enumerate}
This strategy reformulates the PDE as a system of $N$ coupled linear ordinary differential equations, each of the form:
$$c_n'(t) = \sum_{m=0}^{N-1}a_{n,m}c_m(t)$$
A system like this can be solved via any numerical ODE solver, such as Runge-Kutta. An implementation involving any number of time steps should be carried out entirely in the Fourier domain, and mapped back via IFFT so as to obtain V at different time coordinates. 

\subsubsection*{The Pseudo-Spectral Method applied to FitzHugh-Nagumo nonlinearity}
In the presence of nonlinear terms, the spectral method is no longer viable. Consider the $V$ terms from the voltage equation from the FitzHugh-Nagumo model
\[
\frac{\partial V}{\partial t} = \frac{D}{d(x)}\frac{\partial}{\partial x}\bigg(\frac{d^2(x)}{r_L(x)}\frac{\partial V}{\partial x}\bigg) + V(V-a)(V-1).
\]
The first term on the right-hand side is readily addressed by the spectral method as described above, but the cubic in $V$ complicates matters. The formulation of a system of $N$ linear ODEs is not possible here, and simply iterating through time with the spectral representation of the system will not adequately account for changes to due to the cubic term. 

The pseudo-spectral method makes a simple alteration to the original method in order to handle these sorts of nonlinearities by assuming that for each time step, the nonlinearity can be considered as approximately linear. After each iteration of the algorithm, one implements IFFT of the data so that the nonlinearity can be once again computed in real space. The FFT of the right-hand side with this updated nonlinearity is then computed, and another iteration of the algorithm can be carried out. That is to say that each iteration of the ODE algorithm comprises the following steps:
\begin{enumerate}
\item Take the IFFT of the current $V$ Fourier coefficients
\item Compute the nonlinear (in this case, cubic) term in real space
\item Take the FFT of the new linear term
\item Compute the linear term in the same method as described in the section on the spectral method
\item Add the linear and nonlinear Fourier coefficients and then iterate the ODE algorithm
\end{enumerate}
The extension of this method to a system of PDEs is as one would expect: apply this same FFT, IFFT and ODE algorithm to multiple discrete-space functions with the same spatial discretization, so as to convert the system of $n$ PDEs of length $N$ discrete-space signals into a system of $nN$ ODEs. 
\subsection*{Application to swollen axons}
Recall that the system of PDEs in question is given by
\begin{align*}
	\frac{\partial V}{\partial t} &= \frac{D}{d(x)}\frac{\partial}{\partial x}\bigg(\frac{d^2(x)}{r_L(x)}\frac{\partial V}{\partial x}\bigg) + V(V-a)(V-1)-R \\
	\frac{\partial R}{\partial t} &= bV-cR
\end{align*}
Assume that $a$, $b$, $c$, $D$, $d(x)$, $r_L(x)$ are provided, as well as initial conditions for $V$ and $R$. The system naturally has some finite length $L$, so take $\vec{x}$ to be a vector of $N$ evenly spaced points between $-L/2$ and $L/2$. $N$ is normally chosen to be some power of 2, as it will also be used as the number of Fourier modes for each discrete-space function. Take $d(x)$, $r_L(x)$, $V(x,0)$ and $R(x,0)$ at the points in $\vec{x}$ to get length $N$ vectors $d$,$r$,$V(0)$ and $R(0)$. \\ \\
\noindent Recall that the Fourier basis $\{\psi_n\}_{n=-\frac{N}{2}}^{\frac{N}{2}-1}$ is given by $\psi_n = e^{ik_nx}$, where $k_n = \frac{2\pi n}{L}$. This gives a length $N$ vector $\vec{k} = \begin{pmatrix}k_0\\k_1\\\vdots\\k_{(N/2) - 1}\\k_{-N/2}\\k_{(-N/2)+1}\\\vdots\\k_{-1}\end{pmatrix}$. \\ \\
\noindent One proceeds with numerical solution of the system of differential equations. FFT is applied to both $V(0)$ and $R(0)$, and these two vectors are concatenated to obtain a vector of length $2N$. 
\noindent This amounts to the following sequence of operations at each time step:
\begin{enumerate}
\item Extract $\hat{V}(t)$ and $\hat{R}(t)$ from the length $2N$ input vector
\item Take the IFFT of $\hat{V}(t)$ to obtain $V(t)$
\item Use this real-space $V(t)$ to compute $V(V-1)(V-a)$
\item Take the FFT of this nonlinear term and save it for later use
\item Begin to compute the linear term by first taking the point-wise product of $i\vec{k}$ and $\hat{V}(t)$
\item Take the IFFT of this expression and multiply it pointwise by $d^2/r$
\item Take the FFT of this expression and multiply it point-wise by $i\vec{k}$
\item Take the IFFT of this expression of this expression and multiply it by $D/d$
\item Take the FFT of this expression, add the nonlinear term and subtract $\hat{R}(t)$ to obtain the first $N$ entries of the output
\item Take $b\hat{V}(t) - c\hat{R}(t)$ to obtain the last $N$ entries of the output
\item Return these two expressions concatenated as a single $2N$-length vector
\end{enumerate}
In line with the strategy outlined earlier, the nonlinearity is recomputed during each iteration. The output of the ODE solver after $n$ time steps will be an $n\times2N$ matrix. From this, one extracts the data for $V$ and $R$ accordingly, via IFFT.

\subsection*{Damaged Axon Prediction Algorithm}
\subsubsection*{Problem Description}
In considering the activity of networks of spiking neurons, the signals of interest are voltage spikes traveling across axons between neurons. As spikes are of approximately fixed amplitude and duration, the simulation of spiking networks commonly uses a binary representation of spikes: at any given time, a neuron may either emit a spike or remain silent, and may in turn receive spikes from neurons to which it is input-connected; this setting presents no ambiguity or variation to the meaning of the term `spike'. 

On the other hand, the full simulation of an axon impacted by axonal swelling involves a variable-width adaptation of the active cable equation, in which the voltage, time, and space are all continuous. In a damaged axon, spikes that begin at one end of the axon may be transmitted (i.e., they may propagate faithfully to the other end of the axon), or they may be lost. The mechanism by which this loss occurs is nonlinear and dependent in part on the proximity of a spike to other spikes traveling across the axon. 

The necessity to implement a simulation of the variable-width active cable at each axon is computationally taxing and renders implementations for even small networks unfeasible. At the same time, there is a natural interpretation of this lossy transmission in terms of binary coding, wherein the loss of a spike amounts to a single bit flip in the code sequence. As we wish to observe the impact of axonal swelling as described by this active cable model on large networks, we propose the following efficient shortcut.

\subsubsection*{Feature Reduction}
The assumption of a single damage type allows for a straightforward feature reduction of the full axon simulation.
While the active cable model uses continuous time, space and voltage, our spiking neural network simulations use discretized versions of these quantities at each axon. More specifically, if Neuron A emits spikes received by Neuron B, then there are discrete time bins in which spikes may be emitted by neuron A. A spike from Neuron A may arrive at Neuron B at the time step after it has been emitted, so propagation is assumed to be instantaneous. 
Accordingly, we proceed with binary models of space and voltage (i.e., spike vs. no spike), as well as discrete time. Altogether this yields two binary sequences, one for neuron A (input), and the other for neuron B (output).

If an axon is of length $L$, and we simulate over a time interval of duration $T$, the variable-width cable axon simulation occurs in the space-time-voltage space $A = [0,L] \times [0,T] \times \mathbb{R}$. Meanwhile, $N$ time steps of a spiking neural network simulation at one axon occur in the space-time-voltage space $S_N = \{A,B\} \times \{1,2,\ldots,N\} \times \{0,1\}$. It is clear from this alone that the information contained in the active cable model is mostly superfluous for our application.

We can map results in $A$ to $S_N$ by first considering space only at $x=0$ and $x=L$, which map to $A$ and $B$ in the first coordinate of $S_N$ respectively. Time can be binned at arbitrary resolution, and we choose to always bin at the refractory period of the neuron. This is the largest possible bin in which the neuron cannot spike more than once, and so it is possible to classify each bin as containing either one spike or none at all.

It is important to note here that this time binning is solely for the purpose of learning axon functionality, and the update rate for the dynamics of the neurons in our spiking neural network is actually ten times faster. As a result, there are actually two time bin sizes at work in the simulation, but it is the larger one (in which each bin is one refractory period) which is of interest here.

\subsubsection*{Training Set}
To speed up the process of simulating a damaged axon, we look to perform all computations in the feature space $S_N$, while faithfully recreating what would have occurred if the more expensive simulations had been run in $A$. The $S_N$ space is significantly more tractable than $A$ for any choice of $N$, but we cannot feasibly learn the behavior of a damaged axon for \textit{all} $N$. To deal with the problem of arbitrarily large spike trains, we use the fact that the deletion of a spike is caused by spikes near, in time, to it. By ``near'', we mean within a sufficiently large window of time bins around it. To illustrate the necessity of a window larger than $1$ refractory period, there are damage architectures in which two spikes fired one refractory period apart are not deleted, but a third spike fired one refractory period after the second would be deleted. 

Although all spikes have the potential to impact all other spikes, we assume a local impact radius of $M$, where a spike in bin $i$ may be affected by a spike in bin $i+M-1$, bu not in bin $i+M$. This choice is informed simply by observing a few outputs of the active cable simulation and making an educated guess. We pick $M=9$ for our experiments.

Thus, despite the fact that our simulation must deal with spike trains in $S_N$, we can train our system off of spike trains in the smaller $S_M$, and use the behavior of the damaged axon on length $N$ spike trains to infer the behavior of the axon on length $M$ spike trains.

If $M$ is chosen to be sufficiently small, one can observe the effect of the active cable model of the damaged axon on all length $M$ spike trains. There are $2^M$ such spike trains, so the expense of such a task increases exponentially with $M$. We save these input/output pairs for use in our algorithm for arbitrary $N$.

\subsubsection*{Damaged Axon Prediction Algorithm}

The Damaged Axon Prediction Algorithm (DAPA) uses the known operation of the variable-width active cable damaged axon for spike trains in $S_M$ to infer the its operation for spiketrains in $S_N$, where $N\geq M$. Consider a binary string input to a damaged axon of length $N$, where each bit is indexed $b_1$, $b_2$, ... $b_N$. First, the first $M$ bits ($b_1$ through $b_M$) are considered in isolation. We know from our training data what this input alone would map to, and we maintain this guess in memory. We now have a guess for the output values of these first $M$ bits. 

We then move the window forward by one, observing bits $b_2$ through $b_{M+1}$. We can again use our training data to create a guess for these $M$ bits. For bits $b_2$ through $b_M$, this is the second guess made for their value at the output. This guess is also stored in memory.

We continue moving the window and storing guesses in memory until we have made guesses for the output values of all $N$ bits. While only one guess is made for bit $b_1$ and $b_N$, two guesses are made for $b_2$ and $b_{N-1}$, and up to $M$ guesses are made for any arbitrary bit in the input string. We then make a maximum likelihood estimate -- we sum the guesses for each bit, divide by the number of guesses made for that bit and round to either zero or one. 

In simple terms, DAPA can be described as a maximum likelihood estimate made based on locally known behavior. The complexity of DAPA is notably dependent on the length $N$ of the observed string, and a total of $N-M$ guesses are made.

\subsubsection*{An Example of DAPA}

Consider an input string of length $N=21$, given in binary as 101000010001100010000. Using our own data for all 9-bit strings, we arrive at the following table of guesses:

\bigskip

\begingroup

\setlength{\tabcolsep}{4pt} 
\renewcommand{\arraystretch}{1} 
\tiny
\begin{center}
\begin{tabular}{|c |  c c c c c c c c c c c c c c c c c c c c c|}
\hline
Input & 1 & 0 & 1 & 0 & 0 & 0 & 0 & 1 & 0 & 0 & 0 & 1 & 1 & 0 & 0 & 0 & 1 & 0 & 0 & 0 & 0 \\
\hline
Guess 1 & 1 & 0 & 1 & 0 & 0 & 0 & 0 & 1 & 0 & -- & -- & -- & -- & -- & -- & -- & -- & -- & -- & -- & --\\
\hline
Guess 2 & -- & 0 & 1 & 0 & 0 & 0 & 0 & 1 & 0 & 0 & -- & -- & -- & -- & -- & -- & -- & -- & -- & -- & --\\
\hline
Guess 3 & -- & -- & 1 & 0 & 0 & 0 & 0 & 1 & 0 & 0 & 0 & -- & -- & -- & -- & -- & -- & -- & -- & -- & --\\
\hline
Guess 4 & -- & -- & -- & 0 & 0 & 0 & 0 & 1 & 0 & 0 & 0 & 1 & -- & -- & -- & -- & -- & -- & -- & -- & --\\
\hline 
Guess 5 & -- & -- & -- & -- & 0 & 0 & 0 & 1 & 0 & 0 & 0 & 1 & 0 & -- & -- & -- & -- & -- & -- & -- & --\\
\hline
Guess 6 & -- & -- & -- & -- & -- & 0 & 0 & 1 & 0 & 0 & 0 & 1 & 0 & 0 & -- & -- & -- & -- & -- & -- & --\\
\hline
Guess 7 & -- & -- & -- & -- & -- & -- & 0 & 1 & 0 & 0 & 0 & 1 & 0 & 0 & 0 & -- & -- & -- & -- & -- & --\\
\hline
Guess 8 & -- & -- & -- & -- & -- & -- & -- & 1 & 0 & 0 & 0 & 1 & 0 & 0 & 0 & 0 & -- & -- & -- & -- & --\\
\hline
Guess 8 & -- & -- & -- & -- & -- & -- & -- & -- & 0 & 0 & 0 & 1 & 0 & 0 & 0 & 0 & 1 & -- & -- & -- & --\\
\hline
Guess 9 & -- & -- & -- & -- & -- & -- & -- & -- & -- & 0 & 0 & 1 & 0 & 0 & 0 & 0 & 1 & 0 & -- & -- & --\\
\hline
Guess 10 & -- & -- & -- & -- & -- & -- & -- & -- & -- & -- & 0 & 1 & 0 & 0 & 0 & 0 & 1 & 0 & 0 & -- & --\\
\hline
Guess 11 & -- & -- & -- & -- & -- & -- & -- & -- & -- & -- & -- & 1 & 0 & 0 & 0 & 0 & 1 & 0 & 0 & 0 & --\\
\hline
Guess 12 & -- & -- & -- & -- & -- & -- & -- & -- & -- & -- & -- & -- & 1 & 0 & 0 & 0 & 1 & 0 & 0& 0 & 0\\
\hline
Average & 1 & 0 & 1 & 0 & 0 & 0 & 0 & 1 & 0 & 0 & 0 & 1 & 1/9 & 0 & 0 & 0 & 1 & 0 & 0 & 0 & 0\\
\hline 
Output & 1 & 0 & 1 & 0 & 0 & 0 & 0 & 1 & 0 & 0 & 0 & 1 & 0 & 0 & 0 & 0 & 1 & 0 & 0 & 0 & 0\\
\hline
\end{tabular}
\end{center}

\endgroup

\normalsize
\bigskip
This example shows one spike being deleted. This is corroborated by the active cable equation simulation, which as shown in the Figures 1, 2 and 3 gives the same spike train output -- a single deletion in the second half of the train.

\begin{figure} [h]
	\center{\includegraphics[width = .75\textwidth]{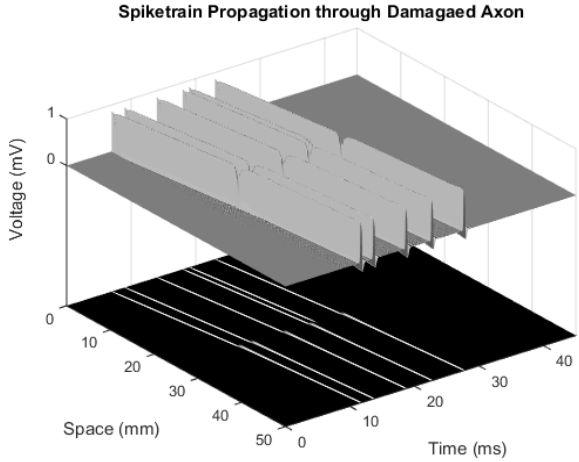}}
	\caption{Propagation of spike train in $A$.}
\end{figure}
\begin{figure} [h]
	\center{\includegraphics[width = .75\textwidth]{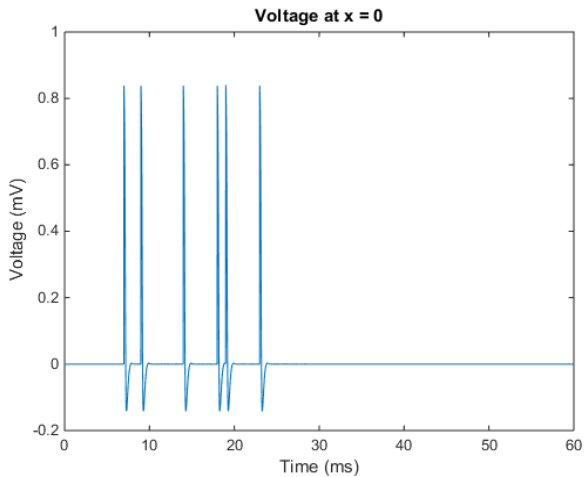}}
	\caption{Input slice, interpreted as the 21-bit train noted in the text.}
\end{figure}
\begin{figure} [h]
	\center{\includegraphics[width = .75\textwidth]{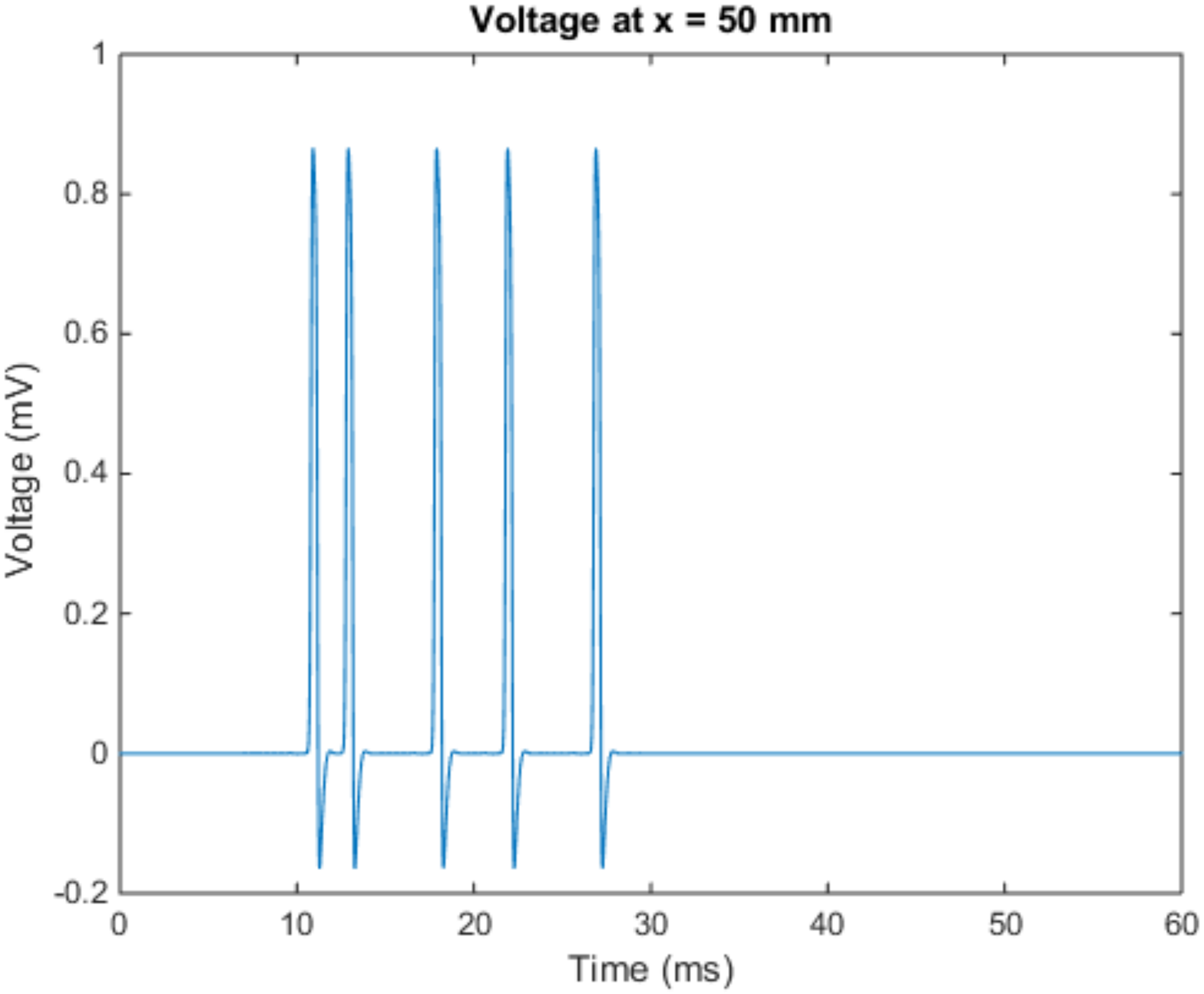}}
	\caption{Output slice, in which one can see that one spike has been deleted in exactly the location predicted by DAPA.}
\end{figure}

\subsubsection*{Observed Performance}

We use two test data sets, each consisting of 1,000 length 99 spike trains generated according to a homogeneous Poisson process. In the first, the Poisson parameter is set to 0.3 and in the second it is set to 0.6. We compare the outputs of the variable-width cable simulation and DAPA bit-wise, and the bit-error rate is simple the total number of differing bits across all test strings divided by the total number of bits in these string (99,000) and multiplying by 100\%. For the first test set, this bit-error rate is exactly 0\%. For the second, this bit-error rate is 0.34\%. 

We also compare the amount of time each simulation takes using the active cable model and DAPA. DAPA is about 4 orders of magnitude faster.

\subsubsection*{Use in Network Simulations}

In network simulations with many time steps, we do not actually look at binary strings whose lengths are the total number of time steps -- this would be inefficient, and impossible, as we do not know which neurons spike when a priori. Instead, we keep a fixed-size buffer containing the last $K$ outputs of each neuron. DAPA can then be performed on this buffer. This also fixes the amount of time taken by DAPA independent of simulation runtime, meaning that a simulation with 100 time steps or 1,000,000 time steps will each have size $K$ buffers used for DAPA. The buffer is always initialized with zeros, signifying that no spikes have fired before the simulation has begun.

\end{document}